\documentclass{emulateapj}
\usepackage{color}

\newcommand{\be}{\begin{equation}}
\newcommand{\ee}{\end{equation}}
\newcommand{\beqn}{\begin{eqnarray}}
\newcommand{\eeqn}{\end{eqnarray}}
\newcommand{\bi}{\begin{itemize}}
\newcommand{\ei}{\end{itemize}}
\def\pomega{\varpi}
\def\ii{{\rm i}}
\def\cc{{\rm c.c.}}

\usepackage{ulem}

\begin{document} 
	
\title{Theory of Secular Chaos and Mercury's Orbit}

\author{Yoram Lithwick\altaffilmark{1} and Yanqin Wu\altaffilmark{2}}
\altaffiltext{1}{Department of Physics and Astronomy, Northwestern University, 2145 Sheridan Rd., Evanston, IL 60208}
 \altaffiltext{2}{Department of Astronomy and Astrophysics, University of Toronto, Toronto, ON M5S 3H4, Canada }

\begin{abstract}

  We
 study the chaotic orbital evolution of planetary systems, focusing on
 {\it secular} (i.e., orbit-averaged) interactions, because these
 often dominate on long timescales.
 We 
  first focus on the evolution of a test particle
  that is forced by
  multiple massive planets.
 To linear order in eccentricity and inclination,
 its
 orbit precesses with constant frequencies.  But nonlinearities
 modify the frequencies, and can shift them into
 and out of secular resonance with the planets' eigenfrequencies, or
 with linear combinations of those frequencies.
 {\it The overlap of
   these nonlinear secular resonances drive secular chaos in planetary
   systems.}
   We quantify the resulting dynamics for the first time
    by calculating the locations and widths of
 nonlinear secular resonances. 
 When results from both analytical calculations
 and numerical integrations are displayed together in a newly
 developed map, the ``map of the mean momenta'' (MMM), the agreement is
 excellent. This map is particularly revealing for non-coplanar
 planetary systems and demonstrates graphically that chaos emerges from
 overlapping secular resonances.  We then apply this
 newfound understanding to Mercury.  Previous numerical simulations
 have established that Mercury's orbit is chaotic, and that Mercury
 might even collide with Venus or the Sun. 
  Guided by intuition from the
   test particle case, we show that Mercury's chaos is primarily
   caused by the overlap between resonances that are nonlinear combinations
   of four modes, the Jupiter-dominated eccentricity mode, the
   Venus-dominated inclination mode and Mercury's free eccentricity
   and inclination. Numerical integration of the Solar system indeed
   confirms that a slew of these resonant angles alternately librate
   and circulate. We are able to calculate the threshold for Mercury
   to become chaotic: Jupiter and Venus must have eccentricity and
   inclination of a few percent. Mercury appears to be perched on the
   threshold for chaos.
 \end{abstract}

\section{Introduction}

The question of the stability of planetary orbits in the Solar system
has a long history, and has attracted the attention of some of the
greatest scientists, including Newton, Laplace, Lagrange, Gauss,
Poincar\'e, Kolmogorov, and Arnol'd.  Newton thought that
interplanetary perturbations are eventually destabilizing, and that
divine intervention is required to restore the planets' orbits to
their rightful places \citep{Laskar96}.  Yet it is only over the last
twenty years that the stability of the Solar system has been
definitively settled, with the aid of computer simulations
\citep{SussmanWisdom88, Laskar89,
  Quinn91,WisdomHolman,Lecar,Laskar09}.  We now know that Newton was
not far off: the Solar system is {\it marginally stable}: it is
unstable, but on a timescale comparable to its age.  In the inner
Solar system, the planets' eccentricities chaotically diffuse on a
billion-year timescale, with the two lightest planets, Mercury and
Mars, experiencing particularly large variations.  In fact, Mercury
has roughly a $1\%$ chance of colliding with Venus or the Sun within
the next five billion years \citep{Laskar09}.  By comparison, the
giant planets in the outer Solar system are well-spaced, and their
orbital elements undergo largely quasiperiodic variations, exhibiting
chaotic diffusion only on extremely long timescales 
\citep{Laskar96,murrayholman99}.

The chaotic dynamics in the inner Solar system is primarily due to
{\it secular}
 interactions 
\citep{Laskar08}.
In general, interplanetary interactions can be decomposed into secular
ones and MMR's (mean motion resonances---not to be confused with
secular resonances).  Secular interactions result from orbit-averaging
the equations of motion.  Since averaging a Keplerian orbit produces
an elliptical ring, secular evolution can be thought of as
interactions between elliptical rings.  Secular timescales are
long---they are longer than the orbital time by at least the ratio of
the star's mass to that of a planet.
By contrast, interactions driven by MMR's depend on orbital phase, and
typically occur on the orbital timescale, or longer if some of the
planets' orbital periods are close to integer ratios.  Intuitively,
one would expect that the dynamics on long timescales can be treated
by averaging over the fast orbital phase---i.e., that they are secular
in nature.  This is true in the inner Solar system.  It is also true
more generally for well-spaced planets that do not happen to lie near
mean motion resonances.\footnote{In the outer Solar system 
the dynamics is not mainly secular
    because the giant planets lie near a number of MMR's, such as the
    5:2 between Jupiter and Saturn (the ``Great Inequality''), and the
    2:1 between Uranus and Neptune.
}

{\it Linear} secular theory has been understood for hundreds of years,
dating back to the famous solution of Laplace and Lagrange
\citep[see][]{MD00}.  To linear order in the planets'
eccentricities and inclinations, secular theory reduces to a simple
eigenvalue problem, with two eigenmodes per planet---one for the
eccentricity degree of freedom, and one for the inclination.  Each
eigenmode has a constant amplitude and a longitude that precesses
uniformly in time.  But linear secular theory
 is clearly incapable of describing the chaotic
orbits  of the Solar system.

Despite the importance of secular chaos, there has been surprisingly
little theoretical understanding of it \citep[see, e.g., the review of
Solar system chaos by][]{Lecar}.  By contrast, chaos due to MMR's is
well-understood, and accounts for the Kirkwood gaps in the asteroid
belt \citep{wisdom83}, and for the very weak chaos of the outer Solar
system planets, which is due to 3-body MMR's \citep{murrayholman99}.
In all cases that have been studied in the Solar system, chaos is
caused by overlapping resonances \citep{Chirikov,Lecar}.  In linear
secular theory, there can be secular resonances.  And it is generally
supposed that the chaos in the inner Solar system is caused by
overlapping secular resonances.  Yet thus far there has been little
quantitative calculation.  To our knowledge, the only previous
theoretical work towards calculating secular chaos was by
\cite{sid90}, who considered the coplanar case, as we describe below
(\S \ref{sec:cop}).

Numerical attempts to identify the mechanism of chaos in the inner
Solar system were made by \cite{laskar90,laskar92} and
\cite{SussmanWisdom}.  These authors found that the angle associated
with the (secular) frequency $(g_{\rm mercury}-g_{\rm
  jupiter})-(s_{\rm mercury}-s_{\rm venus})$ alternately librated and
circulated in their simulations, where $g$ is the apsidal precession
rate, and $s$ is the nodal precession rate (or, to be more precise,
$g$ and $s$ here refer to the frequencies of the normal mode that is
dominated by the corresponding planet).  \cite{laskar92} also found
that two angles associated with Earth and Mars, corresponding to
$2(g_{\rm mars}-g_{\rm earth})-(s_{\rm mars}-s_{\rm earth})$ and
$(g_{\rm mars}-g_{\rm earth})-(s_{\rm mars}-s_{\rm earth})$,
alternately librated, and conjectured that the overlap of those
secular resonances was responsible for chaos.  But, as
\cite{SussmanWisdom} note, Laskar's conjecture is not fully
convincing, because there are too many unrelated angles that
alternately circulate and librate, and it is not clear which are
dynamically important.  Furthermore, only one librating angle has been
identified for Mercury.  Yet chaos requires the overlap of at least
two resonances, so why is Mercury chaotic \citep{Lecar}?
Without a theory for secular chaos,  the dynamics
 remain obscure.  For example,
why does instability in the Solar system occur at such low values
of eccentricity and inclination
 ($\sim$ few percent)? What sets the timescale of the chaos?
 Can secular chaos shape the architecture of the inner Solar system
 \citep{Laskar96}?  And can it shape the architecture of extrasolar
 planetary systems \citep{paper2}?  Without a theory for secular
 chaos, we will be forever at the mercy of computer simulations.

 In this paper, we construct the theory for secular chaos of a test
 particle, and then apply the theory to Mercury.  In \S \ref{sec:sem},
 we present the test particle's equations of motion.  In \S
 \ref{sec:cop}, we describe the coplanar solution, and in \S
 \ref{sec:iej} we generalize to the case when bodies have non-zero
 inclinations.  In \S \ref{sec:realmercury}, we apply the theory to
 N-body simulations of the real Mercury.  We conclude in \S
 \ref{sec:summary}.

\section{Secular Equations of Motion}
\label{sec:sem}

We focus on the secular evolution of a massless test particle that is
orbiting a star in the presence of multiple massive planets, assuming
the planets' orbits are known.

The particle has six orbital elements, $\{a,e,i,\lambda,\pomega,\Omega\}$,
using standard notation \citep{MD00}.  In secular theory, one averages
over $\lambda$.
As a consequence, $a$ is a constant of motion, leaving only
four orbital elements to be considered.
The equations of motion for the particle's eccentricity and
longitude of periapse ($e$ and $\pomega$)
are given by Hamilton's equations for
 the
Poincar\'e canonical variables
$\Gamma\equiv\sqrt{GM_\odot a}\left(1-\sqrt{1-e^2}\right)$
and $\gamma\equiv -\pomega$ \citep{MD00}.
Since $a$ is constant, it is simpler to choose the canonical momentum
to be $\propto \Gamma/\sqrt{GM_\odot a}$, so we introduce the momentum
\beqn
p_e&\equiv& 2\left(1-\sqrt{1-e^2}\right) \\
&=&e^2+{\cal O}(e^4) \ \eeqn and its conjugate co-ordinate to be
$\pomega$.  Although this is a non-canonical transformation from
Poincar\'e's variables, if we simultaneously re-scale the energy by
defining as the Hamiltonian \be H\equiv -{2\over\sqrt{GM_\odot a}}E \
,
\label{eq:hamscal}
\ee
where $E$ is the particle's energy per unit mass\footnote{
The test particle's energy per unit mass $E$ is 
given by equation (\ref{eq:eperunit}) in Appendix A
for the case of  a single external planet.
}
then 
the equations of motion are Hamilton's equations,
\beqn
{d\pomega/dt}&=&\partial H/\partial p_e \label{eq:reom1} \\
{dp_e/dt}&=&-\partial H/\partial \pomega \label{eq:reom2}\ .
\eeqn
Therefore we may consider $(p_e,\pomega)$ to be canonically conjugate.
Similarly, 
 for the inclination and longitude of node $(i,\Omega)$, we
 take the canonical variables 
 to be related to the corresponding Poincar\'e variables in the same way by
defining
\beqn
p_i&\equiv&4\sqrt{1-e^2}\sin^2(i/2)\\
&=&i^2+{\cal O}(e^2i^2,i^4) \ ,
\eeqn
and taking its conjugate co-ordinate to be $\Omega$.
The equations of motion for $(p_e,\pomega,p_i,\Omega)$ are
Hamilton's equations generated by the 
scaled $H$.
These equations are exact as long as $a$ is constant, which
is the case for secular interactions.

An alternative formulation of the equations of motion will also prove
useful.  For an arbitrary Hamiltonian $H(p,q)$, one may define the
{\it complex canonical variable}\footnote{Our definition of the
  complex canonical variable differs from \cite{Ogilvie07} by a minus
  sign, and hence our Hamilton's equation also differs by a minus
  sign.}  $Z\equiv \sqrt{p}e^{iq}$.  As may be directly verified, the
equation of motion for $Z$ is then ${dZ/ dt}=i{\partial H/ \partial
  Z^*} $ where $H(Z,Z^*)=H(p,q)$.  
  This complex equation of motion
simultaneously encodes both of Hamilton's equations.
Since our real canonical variables
are
$(p_e,\pomega,p_i,\Omega)$, we introduce
the complex ones
 \beqn z&\equiv&
\sqrt{p_e}e^{i\pomega}=
\left[2\left(1-\sqrt{1-e^2}\right)\right]^{1/2}e^{i\pomega} \approx
ee^{i\pomega}
 \label{eq:zdef}\\
\zeta&\equiv&\sqrt{p_i}e^{i\Omega}
=2\left(1-e^2\right)^{1/4}\sin(i/2)e^{i\Omega}
\approx ie^{i\Omega}
 \label{eq:zetadef}\ ,
\eeqn
which,
to leading order in $e$ and $i$, are the usual
complex eccentricity and inclination.\footnote{
 The symbol $e$ denotes both the eccentricity
  and the exponential (Euler's constant), and the symbol
 $i$ denotes both the inclination and the imaginary unit.
 There should be no confusion because for the remainder
 of this paper
 we 
  use 
as our dynamical variables either the RCV
or CCV (eqs. [\ref{eq:rcv}]-[\ref{eq:ccv}])
 in lieu of $e$ and $i$.
}
The equations of motion for $(z,\zeta)$ are
\beqn
dz/dt&=&i\partial H/\partial z^*  \\
d\zeta/dt&=&i\partial H/\partial \zeta^*
\eeqn
Throughout this paper, we freely switch between 
the set of real canonical variables and the set of complex ones,
\beqn
&{\rm RCV:}&\  (p_e,\pomega; p_i,\Omega) \label{eq:rcv} \\
&{\rm CCV:}&\  (z, \zeta) \label{eq:ccv} 
\eeqn

Although Hamilton's equations of motion for the RCV and CCV
 are exact, we take an approximate form for the Hamiltonian
 by expanding $H$
 to fourth order in  $e$ and $i$, 
and to leading order in the ratios of semimajor axes, keeping only 
secular terms.
The relevant terms are listed in Table \ref{tab:simp} in Appendix A.
Throughout the bulk of this paper, we focus on the
$c_1, c_2, c_4, c_{11}, c_{12}, c_{14}$, and $c_{15}$
terms in that
table.
As we show in \S \ref{sec:fullfourth},
 the remaining terms in the table have a small effect in the parameter
 regime we focus on.

\section{Coplanar Jupiter and Saturn}
\label{sec:cop}

 In this section, we consider the evolution of a test particle
 in the presence of two massive exterior planets
 (``Jupiter'' and ``Saturn''), with all bodies having zero inclinations.
 We assume that the massive planets' orbits are given by their
 linear Laplace-Lagrange solution, and evolve the test
 particle's equations of motion to leading nonlinear
 order (terms listed in Table \ref{tab:simp} in Appendix  \ref{sec:app1}).
  This case was worked out by
\cite{sid90}.
We describe it here in some detail 
 because it sets the stage for the considerably more complicated
 case with non-zero inclinations (\S \ref{sec:iej}).

\subsection{Jupiter Only}
\label{sec:jonly}

We shall solve the coplanar case with a sequence 
of increasingly complicated sub-cases. 
Consider first the linear secular evolution of 
a test particle perturbed by a circular Jupiter.
From 
Table \ref{tab:simp} in the Appendix, 
the particle's Hamiltonian is
 \be
H(z)=\gamma|z|^2 \ ,
\ee
where the constant
\be
\gamma\equiv {3\over 4}{m_J\over M_\odot}\alpha^3\left(GM_\odot\over a^3  \right)^{1/2}
\ee
is the particle's linear free precession rate induced by Jupiter (eq. [\ref{eq:gammadef}]);
$m_J$ is Jupiter's mass, and $\alpha$ is the ratio of the particle's semimajor axis
to Jupiter's.
The equation of motion is
$dz/dt = i \partial H/\partial z^*=i \gamma z$, with
solution 
 $z={\rm const}\times e^{i \gamma t}$, an orbit 
with constant eccentricity that precesses at frequency $\gamma$.

For the second sub-case, we consider 
the test particle's linear evolution
when Jupiter is assigned a constant eccentricity
$e_J$ and precession rate $g_J$, in which
case the particle's Hamiltonian is
\be
H(z) = \gamma\left(
|z|^2 -(\epsilon_Je^{ig_Jt}z^*+ \cc)
\right) 
\ee
where
$\epsilon_J= {5\over 4}\alpha e_J$; we drop nonlinear terms
(i.e. the fourth-order terms in
Table \ref{tab:simp}).
Of course, if Jupiter were the only massive planet in the system,
it would not precess  ($g_J=0$).
But we consider a finite $g_J$ in anticipation of  what happens
when Saturn is included.

  Hamilton's equation is
 \be
{1\over i \gamma}{dz\over dt}=
z
-\epsilon_Je^{ig_Jt} \ .
\ee
The solution
  is
 a sum of free and forced eccentricities:
 \be
z={\rm const}\times e^{i\gamma t}+
{\epsilon_J\over\Delta} e^{ig_Jt} \ ,
\label{eq:linsol}
\ee
where
\be
\Delta
 \equiv {\gamma-g_J\over\gamma} \ .
 \label{eq:deltadef}
\ee
The forced solution diverges 
at secular resonance ($\gamma=g_J$).
But this divergence is not physical. It is
a consequence of dropping nonlinear terms.

With the nonlinear term included, 
the test particle's Hamiltonian becomes\footnote{
We drop the $c_3$ term  (see Table \ref{tab:simp}) because
it merely alters the coefficient of the $c_1$ term by a small amount.
}
\be
H(z) = \gamma\left(
|z|^2 - {|z|^4\over 4}-(\epsilon_Je^{ig_Jt}z^*+ \cc)
\right) \ , 
\label{eq:hamfund}
\ee
with equation of motion
\be
{1\over i \gamma}{dz\over dt}=
z(
1-{|z|^2\over 2})
-\epsilon_Je^{ig_Jt} \ .
\label{eq:zdot1}
\ee
The nonlinear term reduces the free frequency 
from $\gamma$ to $g\equiv \gamma(1-|z|^2/2)$.
We call $g$ the nonlinear free frequency.  
It can also be introduced 
by first re-writing the Hamiltonian in terms of
the real canonical variables $p_e$ and $\pomega$
and defining
\be
g \equiv {d\pomega \over dt}\big\vert_{\epsilon_J=0}=\gamma(1-p_e/2) \ .
\ee
Because $g$ is a function of
eccentricity, 
secular resonance can occur if the eccentricity is chosen so that
$g\approx g_J$. 
This is a {\it nonlinear secular resonance.}

\begin{figure}
\centerline{\includegraphics[width=0.5\textwidth]{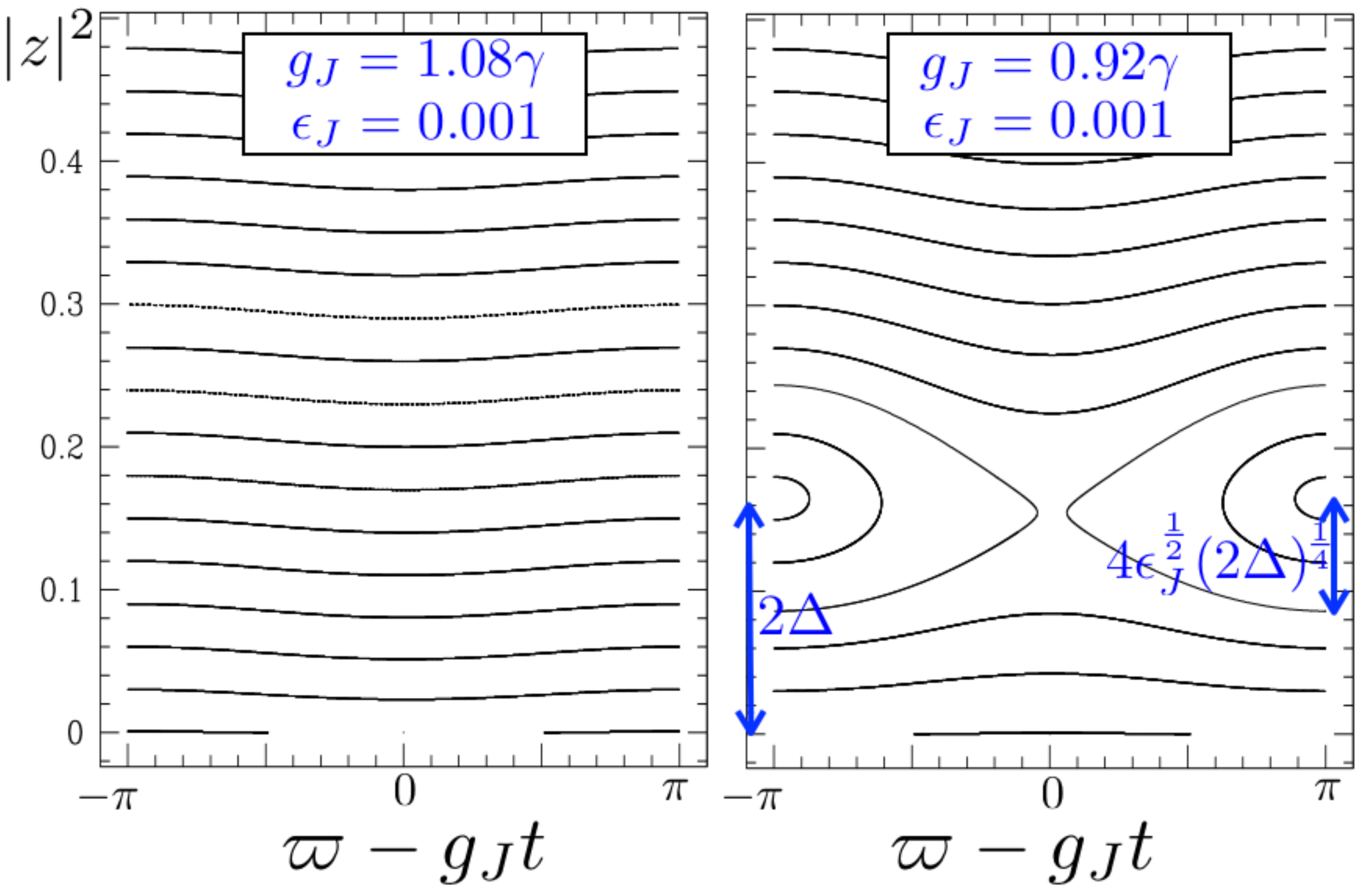}}
\caption{Trajectories of $z$ for a test particle in the Presence of a
  Coplanar Jupiter: \label{fig:single} The curves are solutions of
  Hamiltonian (\ref{eq:hamfund}), i.e. of equation (\ref{eq:zdot1}).
  Equivalently, they are level curves of Hamiltonian (\ref{eq:2nd}).
  The $y$-axis is approximately $e^2$, and the $x$-axis is 
   the phase, 
     modulo $2\pi$. Nonlinear secular resonance occurs for
    the case $g_J < \gamma$.}  
\end{figure}

Solutions of equation (\ref{eq:zdot1}) are shown in Figure
\ref{fig:single} for two cases, $\gamma<g_J$ and $\gamma>g_J$.  When
$\gamma<g_J$ (left panel), the free frequency at small $|z|$
is less than the forcing frequency, $g\vert_{|z|\rightarrow 0}\approx
\gamma < g_J$.  Increasing $|z|$ decreases $g$, moving it further away
from resonance.  Hence the particle's evolution is similar to the
linear case, but with a smaller precession rate.  When $\gamma>g_J$
(right panel), the precession rate at small values of $|z|$ exceeds
$g_J$.  Increasing $|z|$ causes the precession rate $g$ to decrease
until it nearly matches $g_J$.  A resonant island appears, within
which the angle $\pomega-g_Jt$ librates.  Increasing $|z|$ even
further forces the free frequency to be less than $g_J$, 
taking the particle out of secular resonance.

To calculate the width of the nonlinear secular resonance,
   we first write Hamiltonian (\ref{eq:hamfund}) in terms of the real
canonical variables $p_e$ and $\pomega$ (eq. [\ref{eq:zdef}]), 
and then convert the Hamiltonian to a time-independent one by
making the canonical transformation to $(p_e,\pomega_-)$, 
where
\be
\pomega_-
\equiv \pomega-g_Jt \ ,
\ee
yielding the new Hamiltonian 
\be
{1\over\gamma}{H_-}(p_e,\pomega_-)=-{p_e^2\over 4}+
{\Delta}\cdot p_e-2\epsilon_J\sqrt{p_e}\cos\pomega_- \ .
\label{eq:2nd}
\ee
This form for a Hamiltonian has been called
the ``second fundamental model
for resonance,'' the first being the pendulum.
Its
properties have been
extensively catalogued because it also applies to
first-order {\it mean motion} resonances, for which the variables
have a 
different interpretation  \citep{Henrard83,MD00}.
The numerical integrations shown in Figure \ref{fig:single} trace
 level curves
of this Hamiltonian.
For parameters such that the resonant island both exists and
is sufficiently far from $|z|=0$ (as in the right panel of
 Fig. \ref{fig:single}), 
one may approximate the $\sqrt{p_e}$ multiplying the cosine
term as constant, in which case the Hamiltonian is
that of the pendulum.
The
 center of the resonant island is where $(d/dt){\pomega_-}=0$, i.e.
at
\be
p_{e*}=2{\Delta}
\label{eq:pisl}
\ee
in this approximation; this is equivalent to $g=g_J$.  
The width of the island may be found by first completing
the square in the ``kinetic'' part of the Hamiltonian,
i.e. setting $-p_e^2/4+
\Delta\cdot p_e=-{1\over 4}(p_e-p_{e*})^2+$const.
Since ${H_{-}}$ is constant, the half-width of the island is given 
by
\be
 \delta p_{e}=4\left(\epsilon_J\sqrt{p_{e*}}\right)^{1/2}
=4\epsilon_J^{1/2}
(2\Delta)^{1/4} \ .
 \label{eq:dpisl}
 \ee

\subsection{Jupiter and Saturn}
\label{sec:js}
We now add in the effect of a second coplanar planet, Saturn, 
and assume that Jupiter's and Saturn's evolution is described
by their linear Laplace-Lagrange solution.
In that solution,
 Jupiter and Saturn participate in two normal modes, which we
call the Jupiter-dominated and Saturn-dominated modes.
We denote the eigenfrequencies of these two modes
$g_J$ and $g_S$.
Jupiter's (complex) eccentricity is  a sum of two terms,
one for each mode, and may be written as
$e_{J,J}e^{ig_Jt}+e_{J,S}e^{ig_St}$.
Similarly, Saturn's
eccentricity has one term $\propto e^{ig_Jt}$ and another
$\propto e^{ig_St}$. 
 When all four terms are included,
one arrives at the following form for the test particle's Hamiltonian
 (see eq. [\ref{eq:hamfund}]):
\be
H(z) = \gamma\left(
|z|^2 - {1\over 4}|z|^4-(\epsilon_Je^{ig_Jt}z^*+\epsilon_Se^{ig_St}z^*+ \cc) 
\right) \  ,
\label{eq:twoprec}
\ee 
where $\gamma$ is now re-interpreted to represent the test
  particle's precession rate due to both Jupiter and Saturn.  The
  $\epsilon_J$ term is due to the Jupiter-dominated mode, with
  $\epsilon_J$ now a weighted sum of both Saturn's and Jupiter's
  eccentricity within that mode; similarly, $\epsilon_S$ is for the
  Saturn-dominated mode.  For the purposes of this section, rather
  than solving the linear Laplace-Lagrange equations, we consider the
  parameters $\gamma$, $\epsilon_J$, and $\epsilon_S$ to be adjustable
  constants.

 Hamiltonian (\ref{eq:twoprec}) would also describe the test
 particle's evolution if Jupiter's and Saturn's eccentricities and
 precession rates were enforced by hand to be $e_J, e_S, g_J$, and
 $g_S$, with $\epsilon_J={5\over 4}\alpha_Je_J$ and
 $\epsilon_S={5\over 4}\alpha_Se_S$.  This re-interpretation of
 Hamiltonian (\ref{eq:twoprec}), while unphysical as far as Jupiter
 and Saturn are concerned, can be helpful when considering the test
 particle's evolution under their influence.  

\begin{figure}
\centerline{\includegraphics[width=0.49\textwidth]{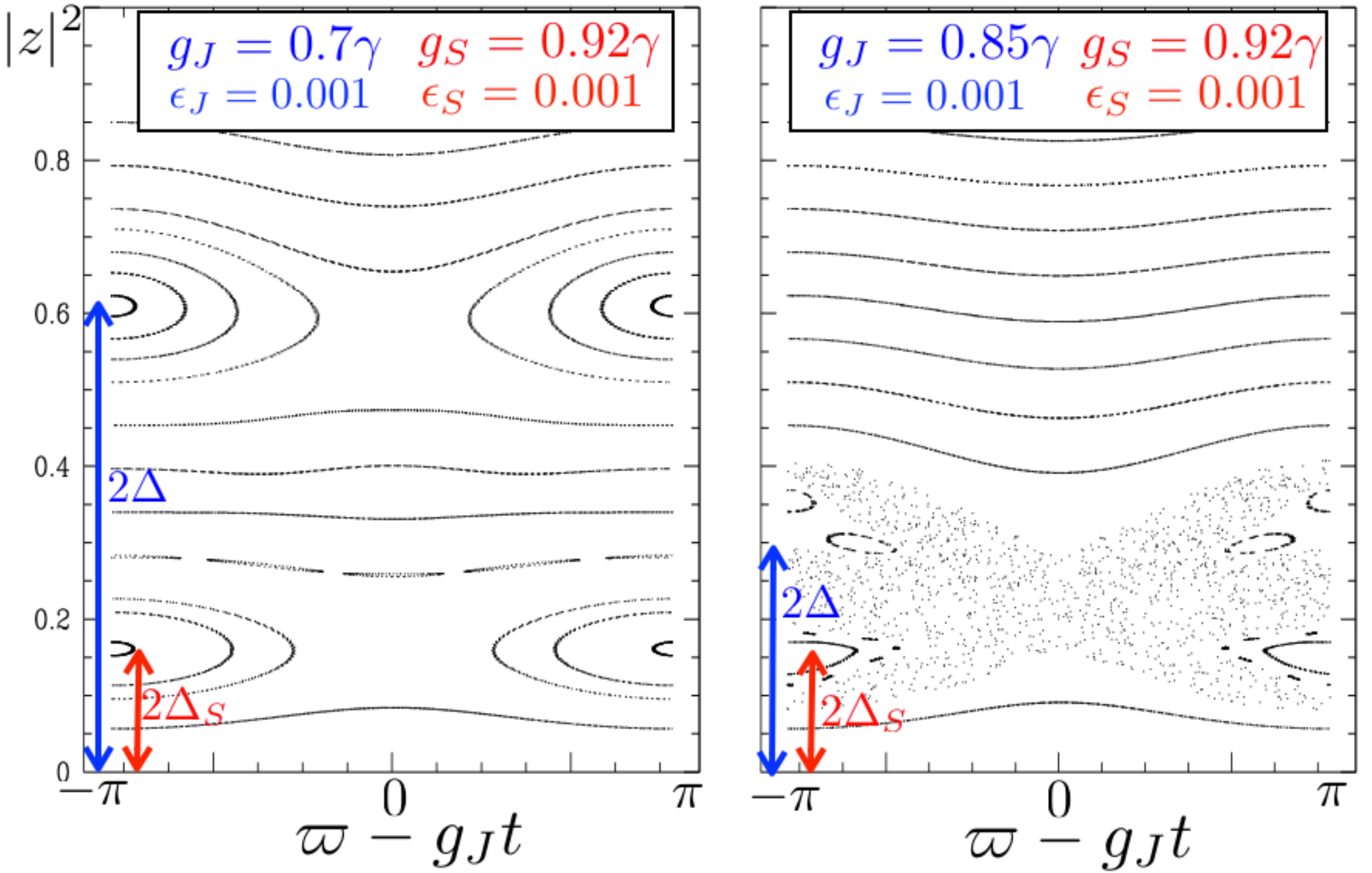}}
\caption{Surfaces of Section for the Coplanar Case ($H$
given by eq. [\ref{eq:twoprec}]):
Section taken at times when  $e^{\ii (g_J-g_S)t}=1$; the
$y$-axis is approximately $e^2$; the $x$-axis is 
the phase, modulo
$2\pi$.
In the left panel, the chosen parameters yield non-overlapping
separatrices, and very little chaos is seen.
In the right panel $g_J$ is slightly larger, yielding overlapping
separatrices and a sea of chaos.
\label{fig:f}
}
\end{figure}

Figure \ref{fig:f} shows results of numerical integrations of Hamilton's equation,
plotted as surfaces of section.
Changing to real canonical variables (eq. [\ref{eq:zdef}]), we see
that Hamiltonian (\ref{eq:twoprec}) has two cosine terms.  The
$\epsilon_J$ cosine term acting alone would yield a resonant island as
long as $\gamma\gtrsim g_J$. Similarly, the $\epsilon_S$ term would
yield an island if $\gamma\gtrsim g_S$.  The location and width of the
islands are quantified in Figure \ref{fig:single}.  When acting
together, there may be two resonant islands.  The left panel of Figure
\ref{fig:f} shows a case when the parameters have been chosen to yield
two non-overlapping islands.  The result is mostly regular motion.
The right panel shows what happens when $g_J$ is increased, so that
Jupiter's
$\Delta$
(eq. [\ref{eq:deltadef}])
 is reduced sufficiently that the islands
overlap: the overlapping islands break up into a sea of chaos.  This
the well-known Chirikov resonance-overlap criterion for chaos
\citep{Chirikov}.  From the widths and locations of the resonances as
displayed in Figures \ref{fig:single} and \ref{fig:f}, one deduces
 that the criterion for chaos 
 is
 $2|g_J-g_S|/\gamma\lesssim 
4 \epsilon_J^{1/2}
(2\Delta)^{1/4}+4\epsilon_S^{1/2}(2\Delta_S)^{1/4}$,  where 
$\Delta_S\equiv (\gamma-g_S)/\gamma$ \citep{sid90}.

It is instructive to consider the form of the surface of section shown
in Figure \ref{fig:f} in more detail.  This reasoning will also be
helpful when we include a second degree of freedom (i.e. inclination)
below.  When the motion of $z$ is regular, it can be written as a
Fourier sum of terms with frequencies equal to the three fundamental
frequencies of the problem (i.e. $g, g_J, g_S$, where $g$ is the
nonlinear free frequency), as well as integer combinations of these
frequencies.  But since only relative frequencies are physically
meaningful, there are really only two fundamental frequencies, which
may be chosen to be $g-g_J$ and $g_S-g_J$, i.e.
 relative to
$g_J$.  In other words, if $z$ is regular, then $\tilde{z}\equiv
ze^{-ig_Jt}$ is a doubly periodic function with periods
 $2\pi/(g-g_J)$ and $2\pi/(g_S-g_J)$.
 In Figure \ref{fig:f}, we choose to plot the
 amplitude versus phase of $\tilde{z}$ whenever the second period
 completed an integer number of cycles.  As long as $z$ is regular,
 the value of $\tilde{z}$ at those times is a singly periodic
 function, and hence appears in the plot as a connected curve.  By
 contrast, when $z$ is chaotic, it appears as scattered points.

\section{An Inclined and Eccentric Jupiter}
\label{sec:iej}
\subsection{Equations of Motion}

\label{sec:iejeom}

In this section, we consider the evolution of a test particle 
that
comes under the influence
of a
single planet (``Jupiter'') that has 
fixed values of eccentricity $e_J$, inclination $i_J$,
 apsidal precession rate $g_J\equiv \dot{\pomega}_J$,
and nodal precession rate 
$s_J\equiv
\dot{\Omega}_J$.

This  is a model for the case when a particle
comes under the influence
 of two planetary Laplace-Lagrange modes, 
one eccentric and one inclined.  
(See the discussion of Hamiltonian [\ref{eq:twoprec}].)
As we show below, in the real Solar system the main modes affecting
Mercury are the Jupiter-dominated eccentricity mode and the
Venus-dominated inclination mode, with the Venus-dominated
eccentricity mode also playing a role.  Therefore for application to
Mercury, $e_J$ and $g_J$ refer to the amplitude and frequency of the
Jupiter-dominated eccentricity mode, while $i_J$ and $s_J$ refer to
those of the {\it Venus}-dominated inclination mode.  Nonetheless, for
the purposes of the present section it is simplest to assume that
Jupiter is the only planet, and that both of its precession rates are
enforced by divine intervention.

We evolve the secular equations for the test particle to leading
nonlinear order.  Because the test particle now has two coupled
degrees of freedom, its evolution is more complicated than before, and
there are many more terms to include in its Hamiltonian.  At first
(\S\S \ref{sec:iejeom}-\ref{sec:sos}), we include only the following
terms from Table \ref{tab:simp}: \beqn {1\over\gamma}H(z,\zeta)=
|z|^2-|\zeta|^2-{|z|^4-|\zeta|^4\over 4}-2|z|^2|\zeta|^2
\nonumber \\
-(\epsilon_J e^{ig_Jt}z^*-i_J e^{is_Jt}\zeta^*+\cc) \ ,
\label{eq:hamsimp}
\eeqn
where $\epsilon_J={5\over 4}\alpha e_J$.
Note that the effect
of Jupiter's eccentricity  is diluted by a factor
 $\sim \alpha\ll 1$, whereas the effect of its inclination
 is undiluted by any such factor.
In \S \ref{sec:fullfourth} we add in all the remaining  
terms from Table
\ref{tab:simp}, and
 show that these additional terms have little effect
in the parameter regime of interest.
The terms in Hamiltonian (\ref{eq:hamsimp}) 
that are second order in eccentricity or inclination
(i.e., the first two terms, and the bracketed terms on the second
line)
are responsible for linear evolution.  
We keep only three nonlinear terms,
$\propto |z|^4, |\zeta|^4$  and $|z|^2|\zeta|^2$.
As we show in this subsection, these are responsible
for nonlinear frequency shifts.  And as we show in subsequent
subsections,  nonlinear frequency shifts are crucial
for resonance overlap and chaos.  Even though the frequency
shifts might be small (second order in eccentricity and inclination), 
they can still be sufficient to shift the frequency into and out of 
secular resonance.
 One of the terms we drop is the Kozai resonance,
i.e.,  the term $c_{26}\times\left(z^{*2}\zeta^2+{\rm c.c.}\right)$ in Table \ref{tab:simp}.
That term has little effect on Mercury's evolution, because
its phase is rapidly varying, and hence the term nearly averages to zero
for parameters similar to Mercury's 
(\S \ref{sec:fullfourth}).
By contrast, the frequency-changing terms can never average
to zero.\footnote{We do drop some  frequency-changing terms, specifically  
ones that are given by  const.$\times |z|^2$, 
and const.$\times |\zeta|^2$, where the constant is ${\cal O}(e^2,i^2)$.
Even though these terms do not average to zero, they merely shift
the linear frequencies, and hence do not
 change the behavior qualitatively.
In the absence of these terms, the 
linear apsidal and nodal frequencies are equal and
opposite; we 
 rectify this shortcoming in our $\kappa$-model Hamiltonian
(eq. [\ref{eq:toy}]).
\label{foot:freq}
}

The equations of motion are
\beqn
{1\over i\gamma}{dz\over dt}&=&z(1-{|z|^2\over 2}-2|\zeta|^2)-\epsilon_Je^{ig_Jt}
\label{eq:zdot}
\\
{1\over i \gamma}{d\zeta\over dt}&=&\zeta(-1+{|\zeta|^2\over 2}-2|z|^2)+i_Je^{is_Jt}  \ .
\label{eq:zetadot}
\eeqn

Expressed in terms of the real canonical variables
 (eq. [\ref{eq:rcv}]), Hamiltonian (\ref{eq:hamsimp})
has two cosine terms (i.e., primary resonances), which become
important when their arguments vary slowly.
 We first focus on the term
  $\propto\cos(\pomega-g_Jt)$.
 Defining the free nonlinear apsidal frequency
\beqn
g \equiv {d\pomega \over dt}\big\vert_{\epsilon_J=i_J=0}
=
\gamma\left(
1
-{1\over 2}{p_e}
-2p_i \right) \label{eq:gdef}   \ , 
\eeqn
one would expect this  resonance to be important near
where
\be
g(p_e,p_i)=g_J  ,
\ee
as may also be inferred by an inspection of equation (\ref{eq:zdot}).
To nonlinear order, 
$g$ is a function of eccentricity and inclination, because of the
nonlinear frequency-changing terms included in Hamiltonian (\ref{eq:hamsimp}).
Hence by varying $e$ and $i$, one can alter the free precession frequency
and bring it into secular resonance. 
In the $p_e$-$p_i$ plane, 
 the resonance traces out a one dimensional curve---or, in fact, a
  straight line to leading nonlinear
order.

The second cosine term in the above Hamiltonian,
 $\cos(\Omega-s_Jt)$, behaves similarly.
Defining 
\beqn
s\equiv
{d\Omega\over dt}\big\vert_{\epsilon_J=i_J=0}
=
\gamma\left(
-1+{1\over 2}p_i -2p_e
\right) \ ,  \label{eq:sdef}
\eeqn
one would expect it to be important near where
\be
s(p_e,p_i)=s_J  .
\ee
We shall make these considerations more precise in 
\S \ref{sec:theory}, where we also work out the
resonant widths, 
 and show that in addition
to the primary resonances are a multitude of secondary
resonances.

\subsection{Simulations: Maps of the Mean Momenta}
\label{sec:sim}

\begin{figure}
\centerline{\includegraphics[width=0.5\textwidth]{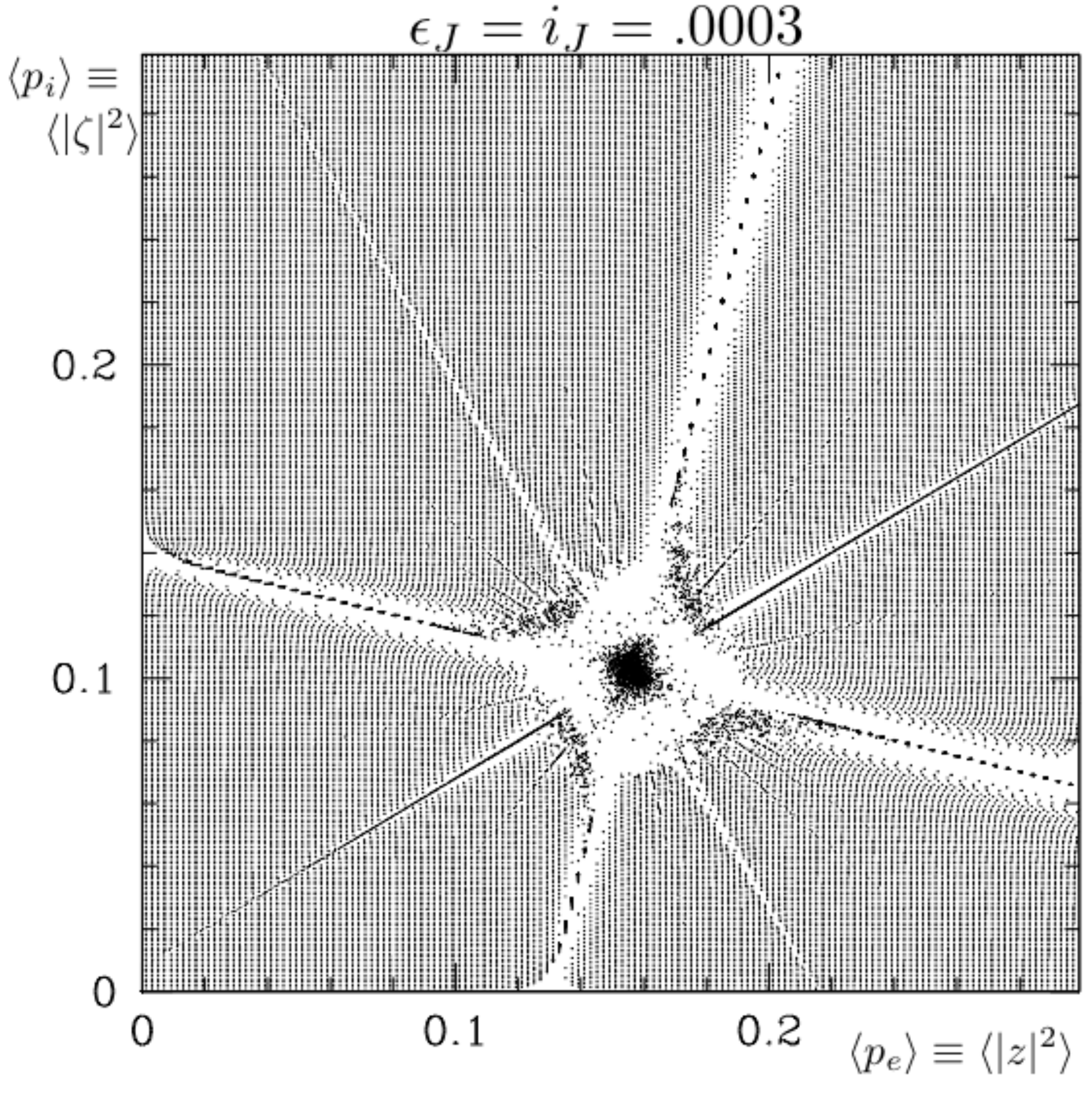}}
\caption{Map of the Mean Momenta (MMM) with low excitation: Each point
  is the result of a single integration of Hamiltonian
  (\ref{eq:hamsimp}), with parameters $\epsilon_J=i_J=.0003$,
  $g_J=0.72\gamma$, and $s_J=-1.26\gamma$.  The $y$-axis is the
  time-averaged $|\zeta|^2\approx i^2$, and the $x$-axis is the
  time-averaged $|z|^2\approx e^2$.  The initial conditions are on a
uniform  grid in $|\zeta|^2$ and $|z|^2$.  Resonant bands are clearly
  visible, as is the zone of chaos where the bands overlap. 
    Because of the averaging, chaotic orbits that explore the full
    extent of the overlap zone give rise to points at the center of
    the zone.  Chaotic orbits that only partially explore the overlap
    zone (as is true of the real Mercury today; see below) give rise
    to chaotic points at the edges of the overlap zone.  The orbits
  are very regular at small $p_e$, $p_i$; for comparison, the real
  Mercury currently has $\langle p_e\rangle \sim .05$ and $\langle
  p_i\rangle \sim .02$.
  Its motion would be regular under this weak forcing.
\label{fig:low}
}
\end{figure}

We run suites of simulations of Hamiltonian (\ref{eq:hamsimp}),
  i.e., of equations (\ref{eq:zdot})-(\ref{eq:zetadot}).  There are
  only five parameters, $\gamma$, $g_J$, $s_J$, $\epsilon_J$, and
  $i_J$.  The linear frequency $\gamma$ sets the overall timescale,
  and can be scaled out.  We choose $g_J/\gamma=0.72$ and
  $s_J/\gamma=-1.26$, since these are close to the true values in the
  Solar system for the Jupiter-dominated eccentricity mode (relative
  to Mercury's free precession frequency), and for the Venus-dominated
  inclination mode (see \S \ref{sec:realmercury}).  We also set the
  excitation amplitudes equal to each other, $\epsilon_J=i_J$, and
  present sequences of simulations with various amplitudes.  The true
  value in the Solar system for the corresponding modes is, very
  roughly, $\epsilon_J\sim i_J\sim 0.01$ (see \S \ref{sec:realmercury}
  for more precise values).

 The dynamics is that of two nonlinearly 
 coupled harmonic
    oscillators, each of which is also nonlinear and
    is forced periodically.
     We have attempted many
    different methods for visualizing the integration results, 
    such as  using catalogs of surfaces of section or Fourier transforms.
    However, most methods were too complicated, and 
    obscured the underlying simplicity 
    of the dynamics, i.e. that it
     is the overlapping
    of resonances that drive chaos.
     In the end, we invented a new method, the Map of
    the Mean Momenta (MMM). This method has many advantages
     over the
    usual surfaces of section.  It is somewhat similar to the frequency
    map analysis of \cite{laskar90} (see below).  

  Figure \ref{fig:low} maps the results from around $50,000$ numerical
  integrations at very small excitation, $\epsilon_J=i_J=.0003$, 
    using the MMM.
  Each point in the plot is the time-averaged value of $|z|^2\approx
  e^2$ and $|\zeta|^2\approx i^2$ from a single simulation, averaged
  over a time span of $3\times 10^4/\gamma$.
  Before taking the time average, we filter with a Hanning filter
  \citep{laskar93}, which leads to a faster convergence of the
  averages (when they do converge).  The simulations were initialized
  with values of $|z|^2$ and $|\zeta|^2$ that were equally spaced on a
  grid, with the spacing in $|z|^2$ twice that in $|\zeta|^2$; the
  initial phases were $\pomega=\pi/2$ and $\Omega=-\pi/2$.

  Three kinds of motion are readily apparent in the MMM: (i) regular
  and non-resonant, (ii) regular and resonant, and (iii) chaotic.
  Most of the figure is covered with a 
  regular grid of points
     that nearly traces the initial conditions.  Here, the values
  of $z$ and $\zeta$ remain regular and non-resonant throughout the
  simulation.  A few resonant bands also appear in the figure, where
  the motion is also regular.  Note that if the initial conditions
  span a resonant island, and if the motion remains regular, then the
  time-averaged momenta ($|z|^2$ and $|\zeta|^2$) exhibit a sharp
  discontinuity.  Inside the island, they average to their values near
  the island center, whereas outside the island's separatrix they
  average to a value offset from the center by a finite amount, of
  order the resonance width (or, more accurately, around 1/4 of the
  resonance full-width).  This can be seen clearly in Figure
  \ref{fig:low}, where the regular points at the center of the
  resonant bands represent librating particles.  Also apparent in the
  figure are the chaotic trajectories.  These show up as the cluster
  of irregular points near where resonant bands intersect.  We have
  checked the Lyapunov exponent, as well as surfaces of section (see
  below), to verify that points that appear on the figure to be
  chaotic are truly chaotic.

\begin{figure}
\centerline{\includegraphics[width=0.5\textwidth]{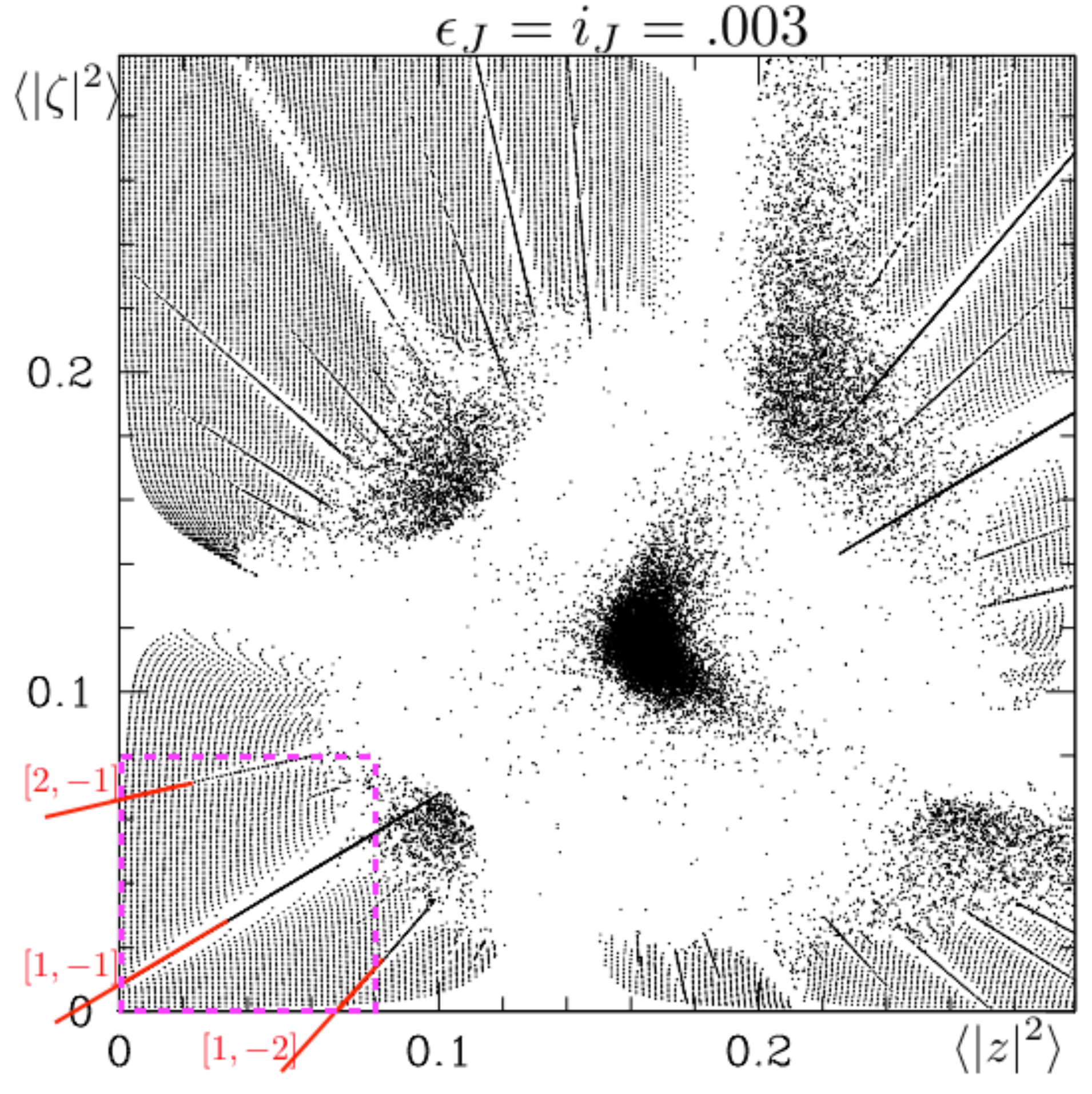}}
\caption{
MMM with Medium Excitation: Similar to Figure \ref{fig:low}, but with $\epsilon_J$ and $i_J$
increased by a factor of 10.  The resonant bands are larger, 
as is the chaotic zone, which has encroached much closer to where the
real Mercury lies,  $\sim (.05,.02)$.
The dashed magenta square is for comparison
with the axes of Figure \ref{fig:pphi}.
\label{fig:ppmed}
}
\end{figure}
\begin{figure}
\centerline{\includegraphics[width=0.5\textwidth]{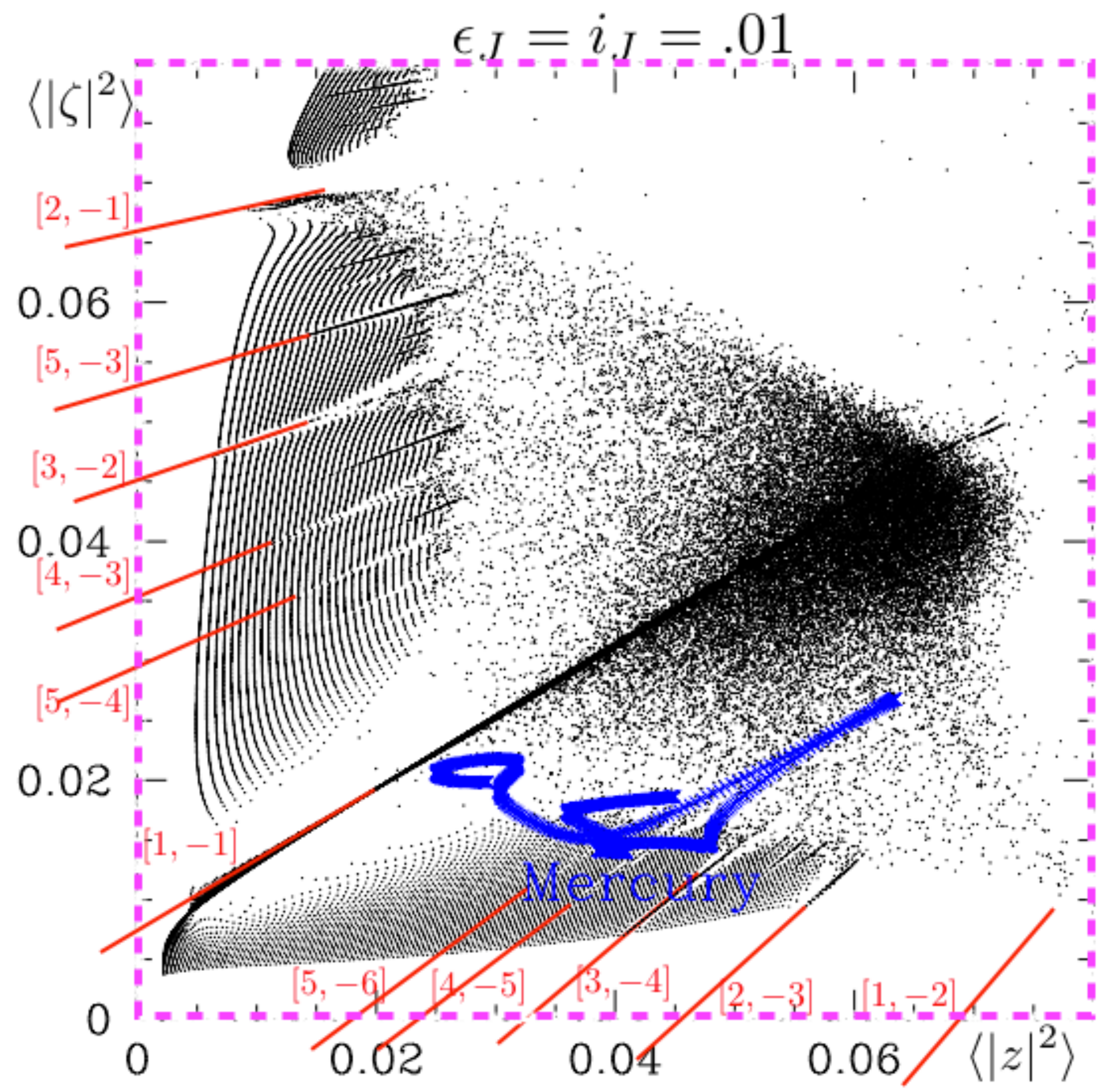}}
\caption{MMM with High Excitation: Similar to Figures
  \ref{fig:low}-\ref{fig:ppmed}, but with $\epsilon_J=i_J=0.01$.  Note
  the expanded scale.  Many high order resonances are visible.  The
  zone of chaos approaches the origin, even though $\epsilon_J$ and
  $i_J$ are $\ll 1$.  The points labelled Mercury are the result of an
  N-body simulation of the full Solar system; each point is Mercury's
  $p_e$ and $p_i$ averaged over a timespan of $100$ Myr, for the first
  $600$ Myr of the simulation shown in Fig. \ref{fig:tmerc}.
  Mercury's true orbit lies near the boundary between regular
  motion and chaos in the MMM of the simplified model.
  Parameters used for this MMM are within $\sim 20\%$ of the the true
  Solar system values.
      The true Solar system is more
    chaotic due to other forcings. 
\label{fig:pphi}
}
\end{figure}

Figures \ref{fig:ppmed}-\ref{fig:pphi} show the MMM for simulations
with the same parameters as in Figure \ref{fig:low}, but with $\epsilon_J$
and $i_J$ increased first to 0.003, and then to 0.01.  With increasing
forcing, the locations of the resonant bands do not change, but they
get wider, and higher order resonances become visible.  As a result,
the zone of chaos where the bands overlap expands.  Surprisingly, even
with the seemingly modest forcing of $\epsilon_J=i_J=0.01$---values
that are comparable to those in the real Solar System (see
below)---the zone of chaos approaches very low values of $e$ and $i$,
and close to the values for the real Mercury.

Our method for displaying results, the MMM, is similar in philosophy
to frequency map analysis \citep[FMA;][]{laskar90}.  But whereas in
FMA one plots the frequencies of the co-ordinates, here we plot the
averages of the momenta.  We have also performed the FMA (not shown);
and when we convert from frequencies to momenta via the inverse of
equations (\ref{eq:gdef}) and (\ref{eq:sdef}), the resulting maps are
almost identical to the MMM.  For the purposes of the present paper,
we prefer the MMM because its axes are approximately $\langle
e^2\rangle$ and $\langle i^2\rangle$, which are simpler to interpret
than the precession frequencies.

\subsection{Theory: Resonance Locations and Widths, and Zone of Chaos}
\label{sec:theory}

To develop understanding of the behavior seen in the MMM's, we first
re-write Hamiltonian (\ref{eq:hamsimp}) in terms of the real canonical
variables (eqs. [\ref{eq:zdef}]-[\ref{eq:zetadef}]), and then
transform from $(p_e,\pomega)$ to $(p_e,\pomega_- \equiv
\pomega-g_Jt)$ and from $(p_i,\Omega)$ to $(p_i,\Omega_-\equiv
\Omega-s_Jt)$, which transforms the Hamiltonian to a time-independent
one: \beqn {1\over\gamma}H_- (p_e,\pomega_-;p_i,\Omega_-)=
{\Delta}\cdot p_e +{\Delta_s}\cdot p_i -{p_e^2-p_i^2\over 4}
\nonumber \\
-2p_ep_i -2\epsilon_J \sqrt{p_e}\cos\pomega_-
+2i_J\sqrt{p_i}\cos\Omega_- \ , \ \ \
\label{eq:hamrot}
\eeqn
where
$\Delta$ is the linear apsidal frequency mismatch
(eq. [\ref{eq:deltadef}]), and \beqn \Delta_s \equiv
{-\gamma-s_J\over\gamma} \eeqn is the mismatch for the nodal
frequencies.
In the absence of the coupling term  ($\propto p_ep_i$) both
degrees of freedom
would evolve independently according to the equations
of the second fundamental model (\S \ref{sec:jonly}).

\begin{figure}
\centerline{\includegraphics[width=0.5\textwidth]{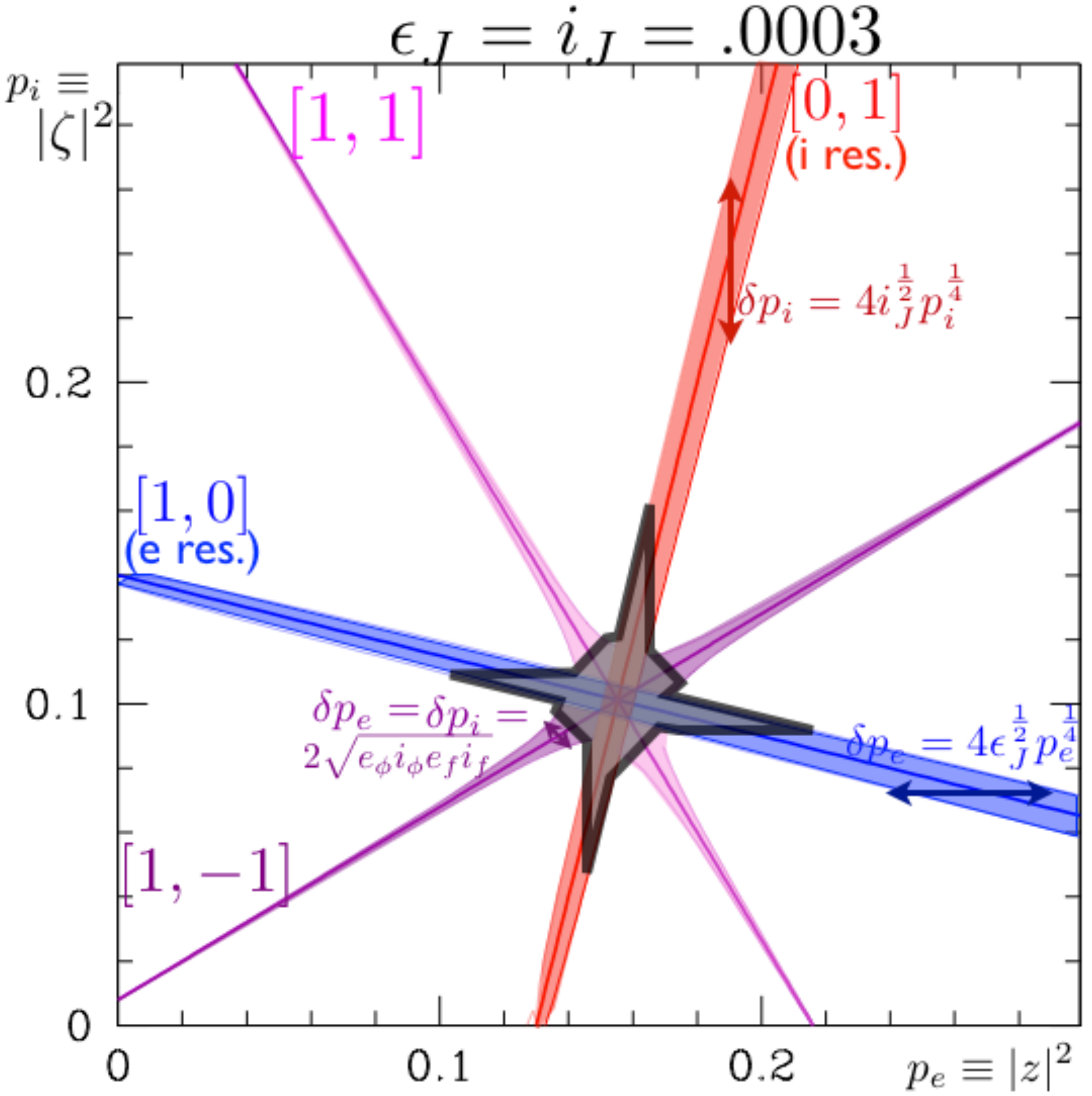}}
\caption{Four strongest resonances and their
widths and region of overlap (analytic calculation):
Parameters are as in Figure \ref{fig:low}.
The center of each $[m,n]$ resonance is the line
determined by $mg_-+ns_-=0$
(eqs. [\ref{eq:freq1}]-[\ref{eq:freq2}]).
The arrows show the half-widths, with the orientation
aligned with the direction of motion in the
corresponding resonances.
The central grey shaded region shows the region of overlapping
separatrices, where chaotic motion is expected.
Comparing with Figure \ref{fig:low} shows
agreement between theory and simulation.
\label{fig:pplot}
}
\end{figure}

In the
  following, we determine the location and width of the 
  two
  primary resonances: the
  eccentricity resonance ([1,0]) and the inclination resonance ([0,1]).
   To do so, we ignore inclination forcing (setting
  $i_J = 0$) when studying the eccentricity resonance, and vice versa.
 The system is 
  trivially integrable if either $i_J=0$ or
$\epsilon_J=0$. In the former case, $p_i$ is constant because the
Hamiltonian does not depend on $\Omega_-$.  Therefore $H_-$ is
equivalent to the coplanar Hamiltonian (eq. [\ref{eq:2nd}]), but with
$\Delta\rightarrow\Delta-2p_i$.  This implies that the center of the
eccentricity resonance is located at 
\be
p_{e*}=2(\Delta-2p_i) \ \ \  \rightarrow [1,0]
\label{eq:10}
\ee
(eq. [\ref{eq:pisl}]) and the island has half-width at fixed $p_i$
given by 
\be
\delta p_{e}=4(\epsilon_J\sqrt{p_{e*}})^{1/2}  \ \ \  \rightarrow [1,0]
\label{eq:w10}
\ee
(eq. [\ref{eq:dpisl}]).  The interior of this island is plotted in
Figure \ref{fig:pplot} as a blue band, with parameter values as chosen
for the simulations of Figure \ref{fig:low}.  Comparing the two
figures shows that the above analytic expressions agree well with the
result of the numerical integrations.  Note that we plot the
half-width in Figure \ref{fig:pplot} rather than the full-width
because the average momentum of an orbit that lies just outside of a
separatrix is approximately half-way to the edge of the resonance.
Figure \ref{fig:pplot} also shows the inclination resonance as a red
band.  Reasoning as before, its center and half-width (at fixed $p_e$)
are
\beqn
p_{i*}=-2(\Delta_s-2p_e)   \ \ \  \rightarrow [0,1]\\
\delta p_{i}=4(i_J\sqrt{p_{i*}})^{1/2}   \ \ \  \rightarrow [0,1]
\label{eq:w01}
\eeqn

In addition to these primary resonances are an infinite number of
secondary ones.  Far from 
resonances, we may
  ignore the cosine forcing terms since they tend to average to
  zero. The test particle's  free precession
  frequencies relative to Jupiter are then
\beqn
g_-&\equiv& g-g_J\equiv {d\pomega_- \over dt}\big\vert_{\epsilon_J=i_J=0}
=
\gamma\left(
\Delta
-{1\over 2}{p_e}
-2p_i \right)  
\label{eq:freq1}\\
s_-&\equiv&s-s_J\equiv
{d\Omega_-\over dt}\big\vert_{\epsilon_J=i_J=0}
=
\gamma\left(
\Delta_s+{1\over 2}p_i -2p_e
\right)
\label{eq:freq2}
\eeqn
Resonances are important near where 
\be
 mg_-+ns_-=0   \ \ \  \rightarrow [m,n]
\label{eq:sig}
\ee for integer pair $[m,n]$.  Therefore the center of each $[m,n]$
resonance traces out a line in the $p_e$-$p_i$ plane. 
For instance, the
centers of the [1,0] and [0,1] resonances are as worked out above, 
and the center of the [1,-1] resonance is the line
\be
p_{e*} - {5\over 3} p_{i*} = {2\over 3} \left(\Delta_s - \Delta\right)
 \ \ \  \rightarrow [1,-1] \ ,
\label{eq:1m1center}
\ee
which passes close to the origin for our choices of $\Delta$ and
$\Delta_s$.
In general, 
the slope of a
resonance line in the $p_e$-$p_i$ plane is $(4n+m)/(n-4m)$,
and
all $[m,n]$ resonant lines intersect at the point in the
$p_e$-$p_i$ plane
 where $g_-=s_-=0$, i.e. at $(p_{e**},p_{i**})$, where 
\beqn
p_{e**}&\equiv&{2\over 17}(\Delta+4\Delta_s) , \ p_{i**}\equiv{2\over
  17}(4\Delta-\Delta_s)
\label{eq:pess}
\eeqn

Although there are an infinite number of secondary resonances, most
are very weak, i.e. their widths are small.
The most prominent resonances in Figure \ref{fig:low} are the 
  primary resonances [1,0] and [0,1], whose widths have been worked
out above.  Next most prominent are the [1, -1] and [1, 1] resonances,
   whose widths
   may be
understood qualitatively as follows (see Appendix B for a quantitative
calculation).  The linear solution for $z$ is a sum of two terms, the
free and forced complex eccentricities (eq. [\ref{eq:linsol}]).
Similarly, to linear order $\zeta$ is a sum of free and forced complex
inclinations.  Therefore to leading nonlinear order, the coupling term
in the Hamiltonian ($p_ep_i=|z|^2|\zeta|^2$) can be written as a sum
of terms, one of which has the form
$z_{\phi}z_f^*\zeta_\phi^*\zeta_f\approx e_\phi e_f i_\phi
i_f\exp(i(g_\phi-g_J- s_\phi+s_J))$, where the subscript $f$ denotes
forced and $\phi$ denotes free.  This term has frequency corresponding
to the [1,-1] resonance; hence the width of this resonance is $\sim
\sqrt{|e_\phi i_\phi e_fi_f|}$.  The [1, 1] resonance behaves
similarly.

The quantitative calculation in the appendix 
shows that, in agreement 
with the above estimate, the
 [1, $\pm 1$] resonances 
have half-widths (eq. [\ref{eq:widapp}])
\be
\delta p_e=\delta p_i=2\sqrt{e_\phi i_\phi e_f i_f} 
 \ \ \  \rightarrow [1,\pm 1]
 \label{eq:w1p1}
\ , 
\label{eq:wid}
\ee after defining the free values as \beqn e_\phi\equiv\sqrt{p_{e*}}
\ , \ \ i_\phi\equiv\sqrt{p_{i*}} \eeqn and the forced values as \beqn
e_f\equiv {\epsilon_J\over |\Delta-p_{e*}/2-2p_{i*}|}
,\ \label{eq:efn} i_f\equiv {i_J\over |\Delta_s+p_{i*}/2-2p_{e*}|}  \ \ \ 
\eeqn 
where the asterisk denotes values at resonant center, and we
neglect here the small difference between the lower-case momenta and
the upper-case ones used in the appendix. 
  As in linear theory, the forced eccentricity 
   scales as the inverse of the frequency
  detuning (eq. [\ref{eq:linsol}]), 
  although now it is the nonlinear frequency detuning
  $g_-$ (eq. [\ref{eq:freq1}]) that is relevant; similarly, $i_f$ is
inversely proportional to $s_-$.

The above widths for the [1,$\pm 1$] resonances are shown in Figure
\ref{fig:pplot}.  The [1, -1] resonance leaves $p_e+p_i$ nearly
constant, which produces trajectories in the $p_e$-$p_i$ plane that
have a slope of -1 (see Appendix B; we again neglect the small
difference between lower- and upper-case momenta).  Therefore for each
value of $(p_{e*},p_{i*})$ at resonant center (i.e., where
$g_--s_-=0$), the upper and lower envelopes of the resonant band are
at $(p_{e*}\pm\delta p_e,p_{i*}\mp \delta p_e)$.  Comparing with
Figure \ref{fig:low} shows excellent agreement between theory and
simulations.  

The central grey region in Figure \ref{fig:pplot} is the overlap zone
of the above four  resonances.  Its shape is
peculiar because each resonance induces a trajectory with a particular
orientation in the $p_e$-$p_i$ plane, and we take the overlapping
region to be wherever one resonance can induce motion into a second
resonance.  The zone of chaos seen in Figure \ref{fig:low} is about
twice as large as that predicted in Figure \ref{fig:pplot}. This
difference is not surprising because we plot the half-widths in Figure
\ref{fig:pplot}, whereas one might expect that chaos would begin where
the full-widths overlap.  Furthermore, we shall show that
higher order combinations of the four strongest resonances
 also play a role in the chaos.
Nonetheless, the grey zone provides a reasonable estimate
of the zone of chaos.

\begin{figure}
\centerline{\includegraphics[width=0.5\textwidth]{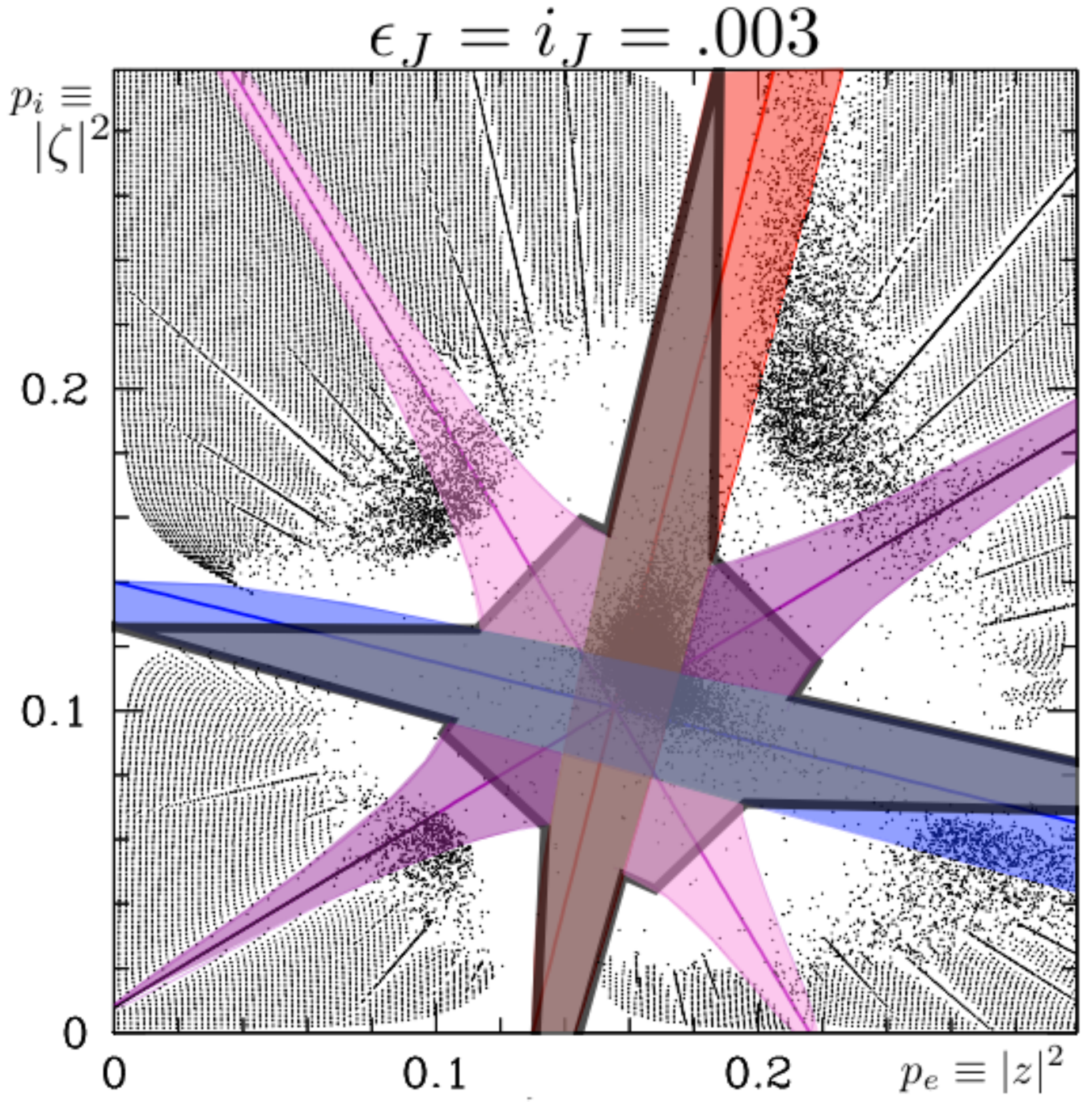}}
\caption{Resonant bands and their region of overlap,
overlayed on MMM:
The colored bands show the theoretically predicted widths
of the four strongest resonances, as in Figure \ref{fig:pplot}.  
The grey region is the zone of resonance overlap for these
four resonances.
Also shown is the  MMM of Figure \ref{fig:ppmed}, 
showing excellent agreement between the theoretical
and numerical resonant widths, and satisfactory agreement for the
zone of chaos.
\label{fig:pplothi}
}
\end{figure}

Figure \ref{fig:pplothi} plots the widths calculated above when the
forcing $\epsilon_J$ and $i_J$
  take on the values of Figure
\ref{fig:ppmed}.  The latter figure is overlayed on Figure
\ref{fig:pplothi} , showing excellent agreement between theory and
simulation in the zones of regular motion.  And, as before, the zone
of chaos from the simulations is around twice as large as the region
where the resonant half-widths overlap.

The dimensions of the zone of chaos may be estimated
analytically.  These
  estimates will be used later to infer the threshold for chaos in
  the Solar system. 
  In Figures \ref{fig:pplot}-\ref{fig:pplothi}, the horizontal and
vertical spikes of the chaotic zone
are due to the
[1, 0] and [0, 1] resonances; 
  the extent of the spikes is simply estimated by the
  resonance width near zone center, \
  \beqn
  \delta
p_{e*}\simeq 4\epsilon_J^{1/2}p_{e**}^{1/4} \ \ \   \rightarrow [1,0]\times [0,1]
\\
\delta p_{i*}\simeq
4i_J^{1/2}p_{i**}^{1/4} \ \ \  \rightarrow [1,0]\times [0,1]
\ ,
\eeqn
 for the horizontal and vertical spikes respectively, 
where $p_{e**}$ and
  $p_{i**}$ are the co-ordinates of the zone center (eq. [\ref{eq:pess}]).  The extent of
the zone caused by the overlap between
the  [1,-1] separatrix with 
the [1,0] 
resonance
   can
  be estimated similarly
  as  
  \be
  \delta p_{e*}=
   k(\epsilon_Ji_J)^{1/4}(p_{e**}p_{i**})^{1/8} \ \ \  \rightarrow [1,-1]\times [1,0]
   \label{eq:over}
   \ee
    where
   $k\simeq (20/17)\sqrt{3/4}$ is an order-unity constant.\footnote{
 Focusing on the region to
  the lower left of the  zone center in Figure  \ref{fig:pplot} or \ref{fig:pplothi},
  the center of the [1,-1] resonance is displaced
  from zone center in the $p_e$-$p_i$ 
  plane by the vector
  $-(1,3/5)x$
     (eq. [\ref{eq:1m1center}]);
      we take $x>0$.
  The orbit given by the half-width of the
  [1,-1] is therefore displaced from zone
  center by the vector
  $-(1,3/5)x+(-1,1)\delta p_e$, 
  where
   $\delta p_e\simeq 2(p_{e**}p_{i**})^{1/4}\sqrt{\epsilon_Ji_J}/(17x/10)$
   (eq. [\ref{eq:wid}]).  Equating that vector to the displacement
   of the center of the [1,0] resonance, i.e. to $(-1,1/4)x'$
   (eq. [\ref{eq:10}])
   yields equation (\ref{eq:over}).
   }
The expression 
   for $\delta p_{i*}$ is the same,
   as are the extents due to the overlap between either 
   [1,1] or [1,-1] with either [1,0] or [0,1], albeit all have different
   order-unity values for $k$.

 It might appear surprising
that the width of the chaotic zone
due to the overlap of a primary resonance 
with a secondary one  (eq. [\ref{eq:over}])
is not much smaller than that due to the overlap
between two primary resonances, despite the fact that
the width of 
a primary resonance is first order in eccentricity
or inclination (eqs. [\ref{eq:w10}]\&[\ref{eq:w01}]), whereas
the width of the [1,-1] is second order (eq. [\ref{eq:w1p1}]).
The reason for this is that the [1,-1] resonance is enhanced by the
denominator that appears in the forced eccentricity and inclination
(eq. [\ref{eq:efn}]). Because of this, the  [1,$\pm 1$] resonances
play
  an important role in
setting the extent of the chaotic zone.
 For example,
  with the forcing frequencies that we have chosen
  for the MMM's (which
  are comparable to those for Mercury), 
  the [1,-1] resonance overlap allows
   the region of chaos to encroach upon the origin
($e\sim 0$, $i\sim 0$) at lower values of the forcing than would have
been expected based solely on the overlap of the [1, 0] and [0, 1]
resonances.   In \S \ref{sec:realmercury}, we shall 
 show explicitly that the
  [1,-1] resonance plays a dominant role in driving chaos for the real
  Mercury. 

One could also  proceed to calculate the widths of higher 
order resonances.
 Extending our reasoning from above, 
  the width of an $[m,n]$ resonance in the $p_e$-$p_i$ plane should
  scale
as 
\be \delta p \sim O(\epsilon^{|m|}i^{|n|}) \ \ \ \rightarrow [m,n] \ ,
\label{eq:mn}
\ee
where $\epsilon$ is comparable to the typical eccentricity, 
and $i$ is comparable to the typical inclination.
However, as we have seen for the $[1,\pm 1]$ resonances, near-resonant
denominators can make the widths significantly larger than
this naive estimate.

\subsection{Surfaces of Section and Libration of Resonant Angles}
\label{sec:sos}

\begin{figure}
\centerline{\includegraphics[width=0.5\textwidth]{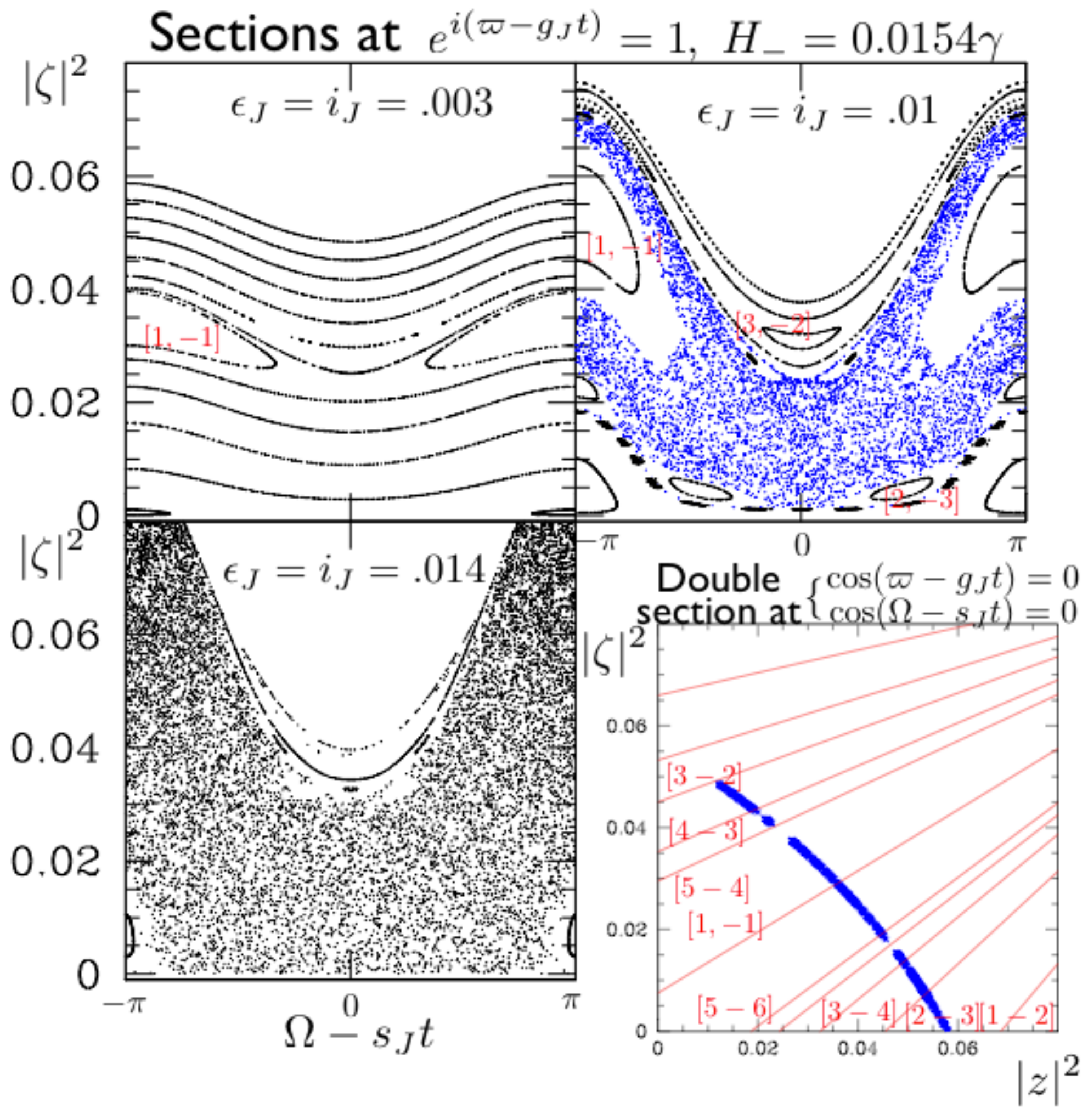}}
\caption{Surfaces of Section
from integrations of Hamiltonian in equation (\ref{eq:hamsimp}):
The parameters are $g_J=0.72\gamma$ and $s_J=-1.26\gamma$, 
and other parameters as shown.  The three upper left panels
show surfaces of section with increasing forcing, showing how
the [1, -1] resonance gets wider, and its separatrix breaks
up into a sea of chaos.  
The (blue) chaotic trajectory in the upper right panel
has, very roughly, parameters comparable to the
real Mercury.
The lower right panel shows a ``double
section'' of this blue chaotic trajectory.
In the double section, it traces out a branch of a hyperbola in
momentum space. Comparing with Figure \ref{fig:pphi} shows 
where this trajectory lies relative to the MMM.
\label{fig:sos}
}
\end{figure}

Here, instead of the global map (MMM), we investigate a
few
  particular trajectories in detail to demonstrate the
   chaotic behavior,
  using both surfaces of section and resonant angles.

Figure \ref{fig:sos} shows a number of surfaces of section from
Hamiltonian (\ref{eq:hamsimp}).  Since the transformed Hamiltonian
(eq. [\ref{eq:hamrot}]) is time independent with two degrees of
freedom, we follow the usual practice of taking a section whenever the
phase of one of the degrees of freedom (here, $\pomega_-$) executes an
integer number of cycles.  At those times, we plot the amplitude
versus phase of the second degree of freedom ($p_i$ vs. $\Omega_-$).
We may also understand this form for the surface of section as
follows.
Hamiltonian (\ref{eq:hamsimp}) has four fundamental frequencies, 
$g_J, s_J, g$, and $s$ where $g$ and $s$ are the nonlinear free
frequencies of $z$ and $\zeta$, respectively.  
Only relative frequencies are physically meaningful, 
and there are three of these, which we may choose
to be $g-g_J$, $s-s_J$, and $g-s$.  But
 Hamiltonian
(\ref{eq:hamsimp}) does not depend on $\pomega-\Omega$, 
and hence the frequency $g-s$ does not enter.\footnote{The
 full fourth order Hamiltonian 
 does depend on $\pomega-\Omega$
 because of the Kozai term;  see 
 Table \ref{tab:simp}.
Therefore one cannot take surfaces of section
of the full fourth order Hamiltonian, but one can 
still plot its MMM (Fig. \ref{fig:pphifull}).
}
Therefore there are two remaining fundamental frequencies,
 as for the
coplanar case  (\S \ref{sec:js}). 
 And, as described there, 
to examine the characteristics of the motion, one may 
take a section whenever the phase 
corresponding to one of the fundamental 
frequencies (here, the phase $\pomega-g_Jt$ which corresponds to $g-g_J$)
executes an integer number of cycles.

The three upper left panels of Figure \ref{fig:sos} show surfaces of
section with values of $g_J$ and $s_J$ as before, and with various
values of $\epsilon_J$ and $i_J$.  All the surfaces of section shown
have the same (constant) value of energy, $H_-=0.0154\gamma$.  To map
out phase space would require many different energy values, but
  for that purpose the MMM is more useful. The
lower right panel in Figure \ref{fig:sos} shows a ``double section''
of the (blue) chaotic trajectory in the upper right panel, i.e. it
shows the two momenta wherever both $\pomega_-$ and $\Omega_-$ have
executed a half-integer number of cycles.  At these times, the cosine
terms in the transformed Hamiltonian vanish, and all trajectories with
a fixed energy $H_-$ fall along the same branch of a hyperbola.  Of
course double sections from all the trajectories shown in Figure
\ref{fig:sos} would lie along the same branch of the hyperbola because
they all have the same value of $H_-$.

The surface of section in the upper left panel has a relatively
low $\epsilon_J$ and $i_J$, and the motion is mostly regular
for the value of energy chosen.  This can also be seen 
in 
the MMM (Fig. \ref{fig:ppmed}) near the relevant 
 hyperbola branch.
The [1, -1] resonance is clearly evident in the top left
panel of Figure \ref{fig:sos}.
Its half-width is  $\delta p_i=0.0055$, as compared 
to  the prediction of $0.006$ from equation (\ref{eq:wid}).
The upper right panel of Figure \ref{fig:sos} shows 
the case with higher forcing.  With this higher forcing, 
the [1, -1] resonance is wider, and the region near its
separatrix has broken up into a wide zone of chaos.
Some of the higher order resonances  are visible 
in this section.  In the lower left panel, the forcing
has been raised further.  Even though $\epsilon_J$
and $i_J$ are still relatively small compared to unity, 
the zone of chaos is vast.

The blue chaotic trajectory in the upper-right panel
of Figure \ref{fig:sos} behaves qualitatively like
the real Mercury, and the parameters are also
similar (\S \ref{sec:realmercury}).  Therefore we investigate it in more detail.
From its surface of section, we see that the separatrix of the
[1, -1] resonance is largely responsible 
for driving the chaos for this orbit, together with
overlapping higher order resonances.
This orbit remains bounded 
by the  [3,-2] and [2,-3] resonances, and hence
can never come under the direct influence of the primary
resonances ([1, 0] and [0,1]).  
The bound on the chaotic zone is a consequence
of using a truncated Hamiltonian that can be written in 
a time-independent form with two degrees of freedom.  
For the full Hamiltonian, one might expect that 
diffusion could act on long timescales (Arnold diffusion),
 ultimately
allowing the trajectory to cross into other regions of
phase space.

\begin{figure}
\centerline{\includegraphics[width=0.5\textwidth]{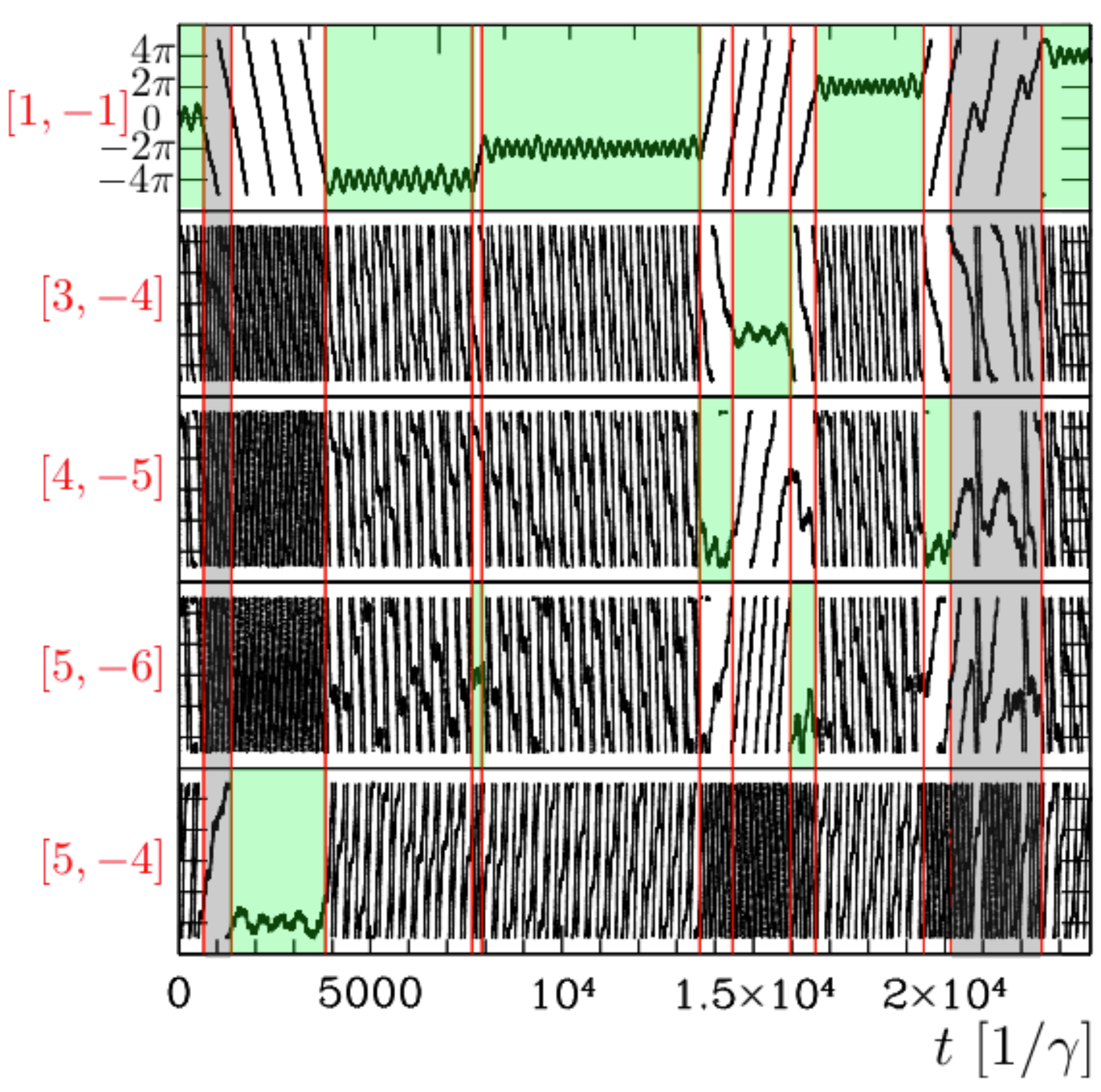}}
\caption{Chain of librating angles: 
Each panel shows 
resonant angles
 $m\pomega_-+n\Omega_-$
 (modulo 10$\pi$)
 with   various values of
 $[m,n]$,
for
  the blue chaotic trajectory of
 Figure \ref{fig:sos}.
The green shaded zones show librating angles.
Different resonant angle combinations librate in turn.
 The first grey strip is when [6,-5]  librates (not shown).
The second grey shaded strip shows a time when 
the
[1,-1] and [4,-5] alternately librate in rapid succession, and
 no other angles are clearly librating.
\label{fig:resang}
}
\end{figure}

Figure \ref{fig:resang} shows explicitly that the chaos
is due to the overlapping of high order resonances.
 The resonant angles $m\pomega_-+n\Omega_-$ are plotted
for various values of $[m,n]$.  Different resonant angles
librate in turn, 
showing that this orbit first comes
under the influence of the [1,-1], then the [5,-4], then 
the [1,-1], etc.

\subsection{Full Fourth Order Hamiltonian}
\label{sec:fullfourth}

\begin{figure}
\centerline{\includegraphics[width=0.5\textwidth]{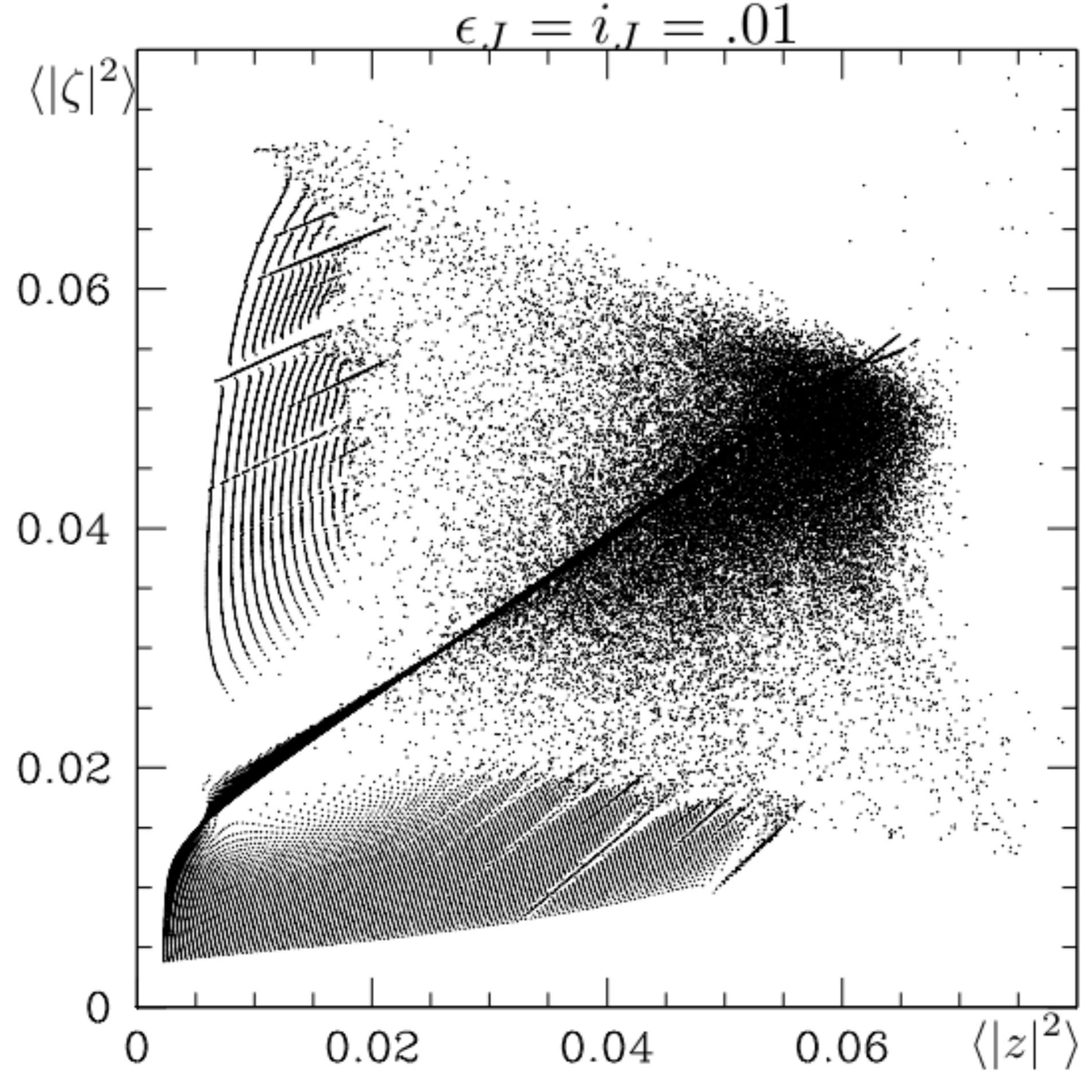}}
\caption{MMM with full fourth order Hamiltonian: 
Similar to Figure \ref{fig:pphi}, 
but integrations have been performed with the full Hamiltonian
(Table \ref{tab:simp} in Appendix A), rather than the truncated
Hamiltonian (\ref{eq:hamsimp}).  
From the fact that the two figures are broadly similar, one can
infer that the terms dropped from Hamiltonian (\ref{eq:hamsimp})
are of small importance in the regime of interest.
\label{fig:pphifull}
}
\end{figure}

Thus far we have focused on the truncated fourth order Hamiltonian
(eq. [\ref{eq:hamsimp}]).  Figure \ref{fig:pphifull} shows
the MMM of the full fourth order Hamiltonian, expanded to leading
order in $\alpha$ (i.e., including all terms in Table \ref{tab:simp}
in Appendix A).  From the similarity of Figure \ref{fig:pphifull} to
the truncated integrations of Figure \ref{fig:pphi}, we conclude that
the terms dropped in the truncated Hamiltonian have little effect on
the dynamics, particularly in the region of small $e$ and $i$ (lower
left corner of the MMM).

The dropped terms have little effect because the only new fundamental
relative frequency they introduce is $g-s$ (see first paragraph of \S
\ref{sec:sos}).  This ``Kozai frequency'' differs significantly from
zero in the domain of Figure \ref{fig:pphifull}, and hence it can only
combine with the other two relative frequencies ($g-g_J$ and $s-s_J$)
to give resonances at high order.  To be quantitative, the resonant
line of the Kozai frequency is at $g-s\approx
\gamma(2+(3/2)p_e-(5/2)p_i)\approx 0$, i.e., it traces the line
$p_i\approx (4+3p_e)/5$.  Hence the Kozai resonance is at much larger
$p_i$ than shown in Figure \ref{fig:pphifull}.

Aside from the Kozai term ($c_{26}$ in Table \ref{tab:simp}), all
other terms dropped from the truncated Hamiltonian depend on Jupiter's
eccentricity or inclination.  These do not introduce new forcing
frequencies because Jupiter's frequencies already appear in the test
particle's orbit at linear order---in its forced $e$ and $i$.  While
the dropped terms do change the amplitudes of the forcing terms, the
change is small as long as Jupiter's $e$ and $i$ is smaller than the
$e$ and $i$ it linearly forces in the test particle, as is true of
Figure \ref{fig:pphifull}.

We suspect that the terms dropped from Hamiltonian (\ref{eq:hamsimp})
are quite often of secondary importance.  This is largely true for the
real Mercury (\S \ref{sec:realmercury}).  And we suspect that it is
true more generally because if secular interactions between two
planets are strong, then the forced $e$'s and $i$'s 
will typically 
(though not always)
be larger
than the forcing ones.
Nonetheless, the dropped terms can be important
in certain circumstances;  for example,
the
Kozai term will play a role if a planet has a high inclination, 
and MMR's will be important for planets whose orbital periods
are near integer ratios.

\subsection{Fourier Transforms}

In \S \ref{sec:realmercury} we shall 
make the connection  to the real Mercury.
For that purpose, it will prove instructive to examine
trajectories in Fourier space. 

For a  more exact comparison to Mercury, we consider here the Hamiltonian
\beqn
{1\over\gamma}H(z,\zeta)=
|z|^2
-\hat{\gamma}|\zeta|^2-{|z|^4-|\zeta|^4\over 4}-2|z|^2|\zeta|^2 
\nonumber \\
-(\epsilon_J e^{ig_Jt}z^*-i_V e^{is_Vt}\zeta^*+\cc) \ ,
\label{eq:toy}
\eeqn which differs from Hamiltonian (\ref{eq:hamsimp}) by the
inclusion of a constant $\hat{\gamma}$ to allow 
the linear 
  apsidal and nodal precession rates to differ from
  each other (see footnote \ref{foot:freq}).
   Note that we also change
notation so that $i_V$ and $s_V$ are the amplitude and precession rate
of the Venus mode.  We focus on a one dimensional family of
systems parameterized by $\kappa$, which
scales all eccentricities and inclinations.  More precisely, in this
``$\kappa$-model'' we choose the parameters
$\epsilon_J=i_V=0.008\kappa$; and initial conditions
$|z|=0.16\kappa$, $|\zeta|=0.07\kappa$, 
 $\pomega=\Omega=\pi/2$.
 The remaining parameters are
 $\hat{\gamma}=0.9$, $g_J=0.72\gamma$,
$s_V=-1.14\gamma$.
With these parameters, the center of the [1,0]
resonance is at $p_{e*}=2(0.28-2p_i)$, as before; 
and the center of the [0,1] resonance is at 
$p_{i*}=-2(0.24-2p_e)$, whereas before the constant
was 0.26 rather than 0.24.  This difference is of little
consequence.
For 
 displaying the results of the integration, we
shall choose
 $\gamma=5.87''/$yr.  
 Our rationale for choosing these particular numerical
 values will be explained in \S \ref{sec:realmercury}.

\begin{figure}
\centerline{\includegraphics[width=0.5\textwidth]{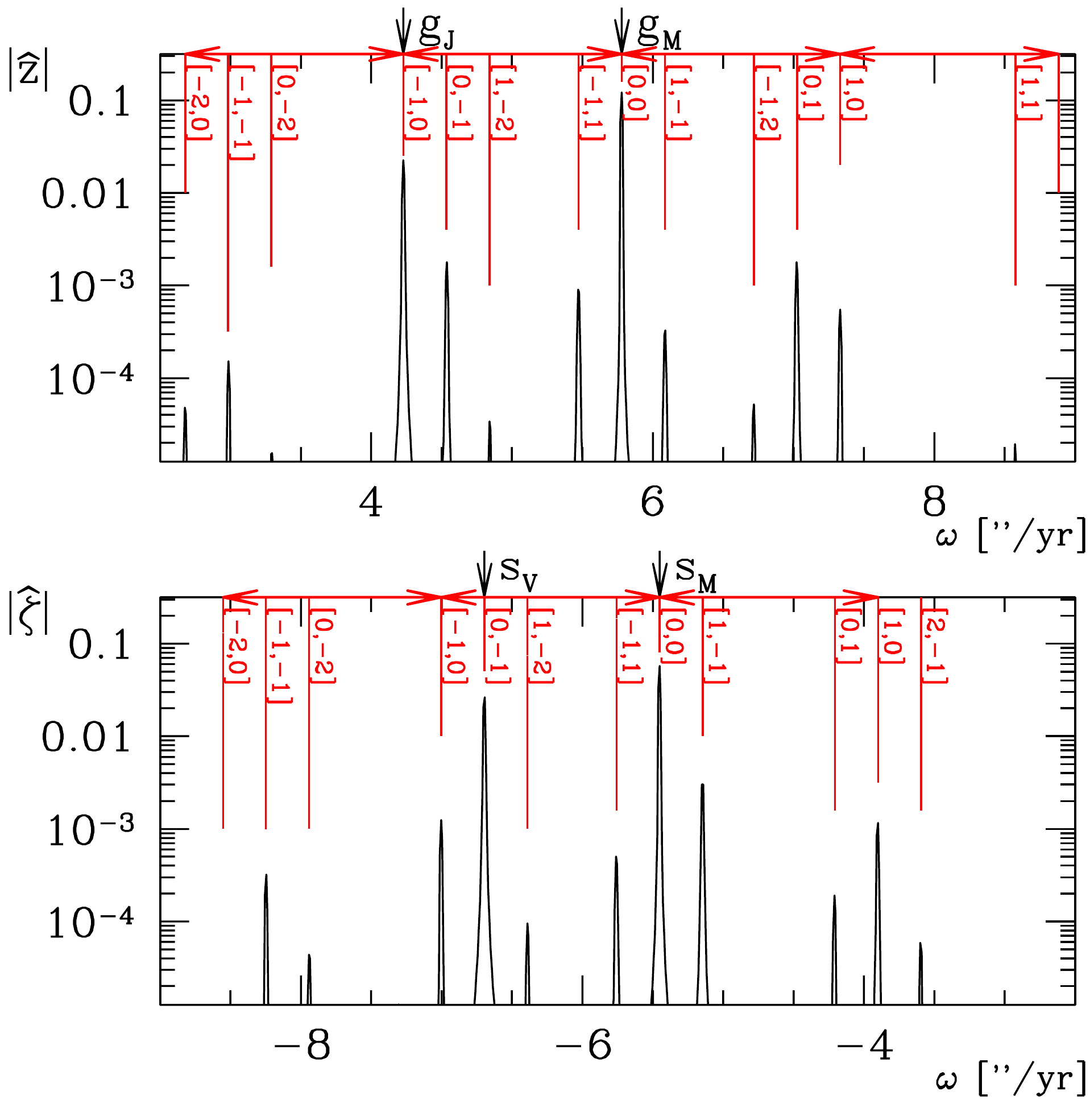}}
\caption{Fourier transforms of the test particle's $z$ and
    $\zeta$ for the $\kappa$-model with $\kappa=0.75$
  (eq.
  [\ref{eq:toy}]):
  With this relatively small value of $\kappa$, the trajectory is
  quasiperiodic, as indicated by narrow spikes in the Fourier
  transform.  The free and forced $z$ are peaks at $g_M$ and $g_J$ in
  the top panel.  The other peaks are due to nonlinear couplings, and
  are at frequencies $g_M+m(g_M-g_J)+n(s_M-s_V)$, labelled $[m,n]$.
  The horizontal red arrows in both panels denote frequencies spaced
  by $g_M-g_J$.  The bottom panel 
   shows that
  $|\hat{\zeta}|$ is similar, with free and forced $\zeta$ at
  frequencies $s_M$ and $s_V$, and nonlinearly generated peaks at
  $s_M+m(g_M-g_J)+n(s_M-s_V)$. 
 \label{fig:toy75}
}
\end{figure}
Figure \ref{fig:toy75} shows the Fourier transforms of $z$ and $\zeta$
for the $\kappa$-model trajectory that has $\kappa=0.75$.  We
normalize the Fourier transform of $z$ as \be
\hat{z}(\omega)\equiv{1\over T}\int_0^T z(t)e^{-i\omega t}dt \ , \ee
and similarly for $\zeta$, where $T$ is the duration of the Fourier
transform, which we choose in the present subsection to be
$T=4400/\gamma$.  With this normalization, if $z$ has constant
amplitude and frequency, i.e. if $z(t)=k_0e^{i\omega_0t}$, then
$|\hat{z}|= |k_0|$ at frequencies close to $\omega_0$.

The trajectory used for Figure \ref{fig:toy75} is quasiperiodic---the
peaks in the Fourier transform are simply spikes, whose widths become
narrower for larger $T$.  We call the two largest peaks in the top
panel the forced and free $z$.  The forced $z$ is at frequency
$g_J=0.72\gamma=4.23''/$yr.  The free $z$ is at frequency
$g_M=5.78''$/yr. Because of nonlinearities, the free frequency ($g_M$)
differs from the linear free frequency ($\gamma$) by a small but
non-negligible amount.  Similarly, in the bottom panel the largest two
peaks are the forced $\zeta$ at frequency $s_V=-1.14\gamma$, and the
free $\zeta$ at frequency $s_M=-5.45''$/yr, which differs from the
linear free frequency $-\hat{\gamma}\gamma$.

In addition to the free and forced $z$ and $\zeta$, there are a
multitude of peaks in Figure \ref{fig:toy75} that are generated by
nonlinear couplings.  The peaks in $\hat{z}$ 
all fall
at frequencies $g_M+m(g_M-g_J)+n(s_M-s_V)$ for integers $m,n$. 
Roughly speaking, the peak amplitudes become smaller for larger values
of $|m|$ and $|n|$.  These amplitudes can be calculated
perturbatively, as is sketched in the following.  As before, we define
the free and forced components as $(z_\phi,z_f,\zeta_\phi,\zeta_f)$,
which have phases that rotate with frequencies $(g_M,g_J,s_M,s_V)$,
respectively.  To leading nonlinear order, the nonlinear terms in the
equation for $dz/dt$ are proportional to
$z|z|^2=(z_\phi+z_f)|z_\phi+z_f|^2$ and
$z|\zeta|^2=(z_\phi+z_f)|\zeta_\phi+\zeta_f|^2$ (eq. [\ref{eq:zdot}]).
These generate six new frequencies in $z$: $(2g_M-g_J), (2g_J-g_M),
g_M\pm (s_M-s_V), g_J\pm (s_M-s_V)$.  Each frequency-generating term
acts as a linear forcing on $z$.  Together with the free and forced
$z$, these account for eight of the peaks marked in the top panel of
Figure \ref{fig:toy75}; specifically, they account for the two highest
peaks in each of the four left-most triplets.  The other peaks are
accounted for by higher order nonlinear terms.  One of these other
peaks---the one labelled [1,-1]---is quite large, even though one
might naively have expected that it would be smaller because it enters
at a higher nonlinear order.  The reason for this is that its forcing
frequency differs from $g_M$ by 
$\sigma \equiv (g_M - g_J) - (s_M - s_V)$
which is quite small.  Hence this near resonance
amplifies the peak by $g_M/\sigma\sim 20$.  We note parenthetically
that the width of the [1,-1] resonance (as described in \S
\ref{sec:theory}) is directly related to the amplitudes of the three
peaks at $g_M$ and $g_M\pm \sigma$.  The Fourier transform of $\zeta$
behaves similarly to that of $z$, with the frequency peaks at
$s_M+m(g_M-g_J)+n(s_M-s_V)$.

\begin{figure}
\centerline{\includegraphics[width=0.5\textwidth]{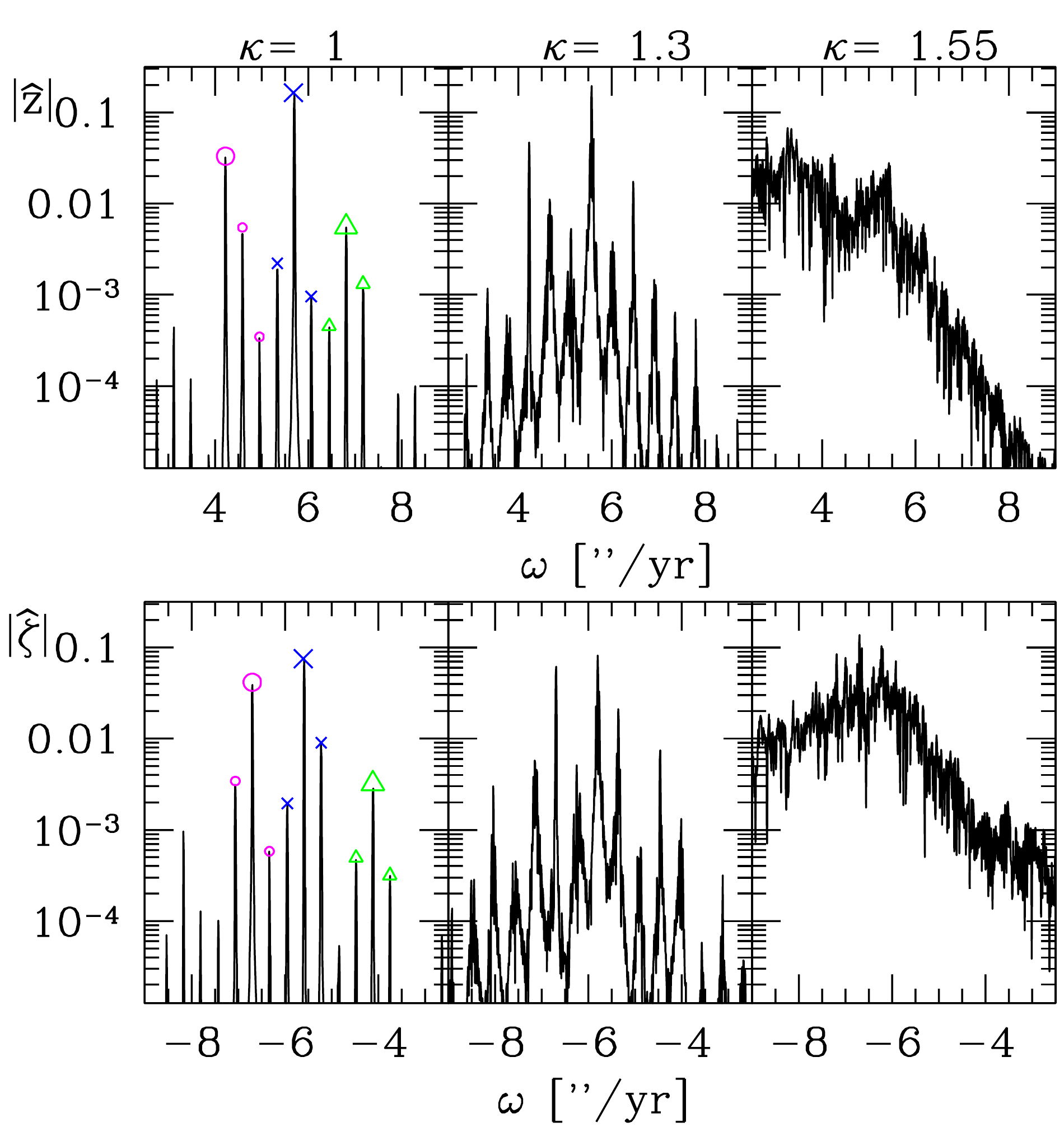}}
\caption{Fourier transforms of results from three 
  $\kappa$-model simulations,
  with $\kappa=1, 1.3, 1.55$. 
   At
  $\kappa=1$ (left-most panels), the motion is still largely
  quasiperiodic.  The amplitudes of the nonlinearly generated peaks
  have risen significantly relative to 
  Fig. \ref{fig:toy75} ($\kappa=0.75$),
   even though the forced and free $z$ and
  $\zeta$ have changed by a modest amount. 
    At
  $\kappa=1.3$, there is weak chaos---the peaks have widened, and
  neighboring peaks overlap.  At $\kappa=1.55$, the trajectory is
  highly chaotic.
\label{fig:toyfftall}
}
\end{figure}

Figure \ref{fig:toyfftall} shows the Fourier transforms for the
$\kappa$-model at higher values of $\kappa$.  The left-most panels
 show the case $\kappa=1$.  The motion is largely
quasiperiodic, but the nonlinearly generated peaks have increased
significantly relative to the $\kappa=0.75$ case.  At $\kappa=1.3$ the
motion is chaotic, and at $\kappa=1.55$ it is highly chaotic.

\section{Mercury}
\label{sec:realmercury}

We integrate the eight Solar system planets with the SWIFT symplectic
integrator \citep{levisonduncan}, supplemented with a routine for
Mercury's relativistic precession \citep[see][for code
details]{paper2}.  We initialize the planets with their current
 orbits and use their
actual masses.  The integration timestep is 8 days.

One might suspect that 
Mercury's orbital evolution is more complicated than our toy model
 for a variety of reasons: its $e$ and $i$ are not too
small, and hence the fourth order expansion is approximate; it is not
massless, and hence backreacts onto the other planets (especially
Venus); there are seven other planets that do not have constant
orbital elements and frequencies but participate in the overall chaos
of the Solar system; and Mercury can be affected by resonant terms.
Despite these complications, we show that the chaotic behavior of
Mercury is qualitatively similar to the Hamiltonian model.  This is
perhaps not too surprising, since nonlinear dynamics are largely
driven by resonances and their overlap. Hence as long as a model
roughly captures the locations and widths of the principal resonances,
it should produce qualitatively correct behavior.

\subsection{Fourier Transforms}

\begin{figure}
\centerline{\includegraphics[width=0.5\textwidth]{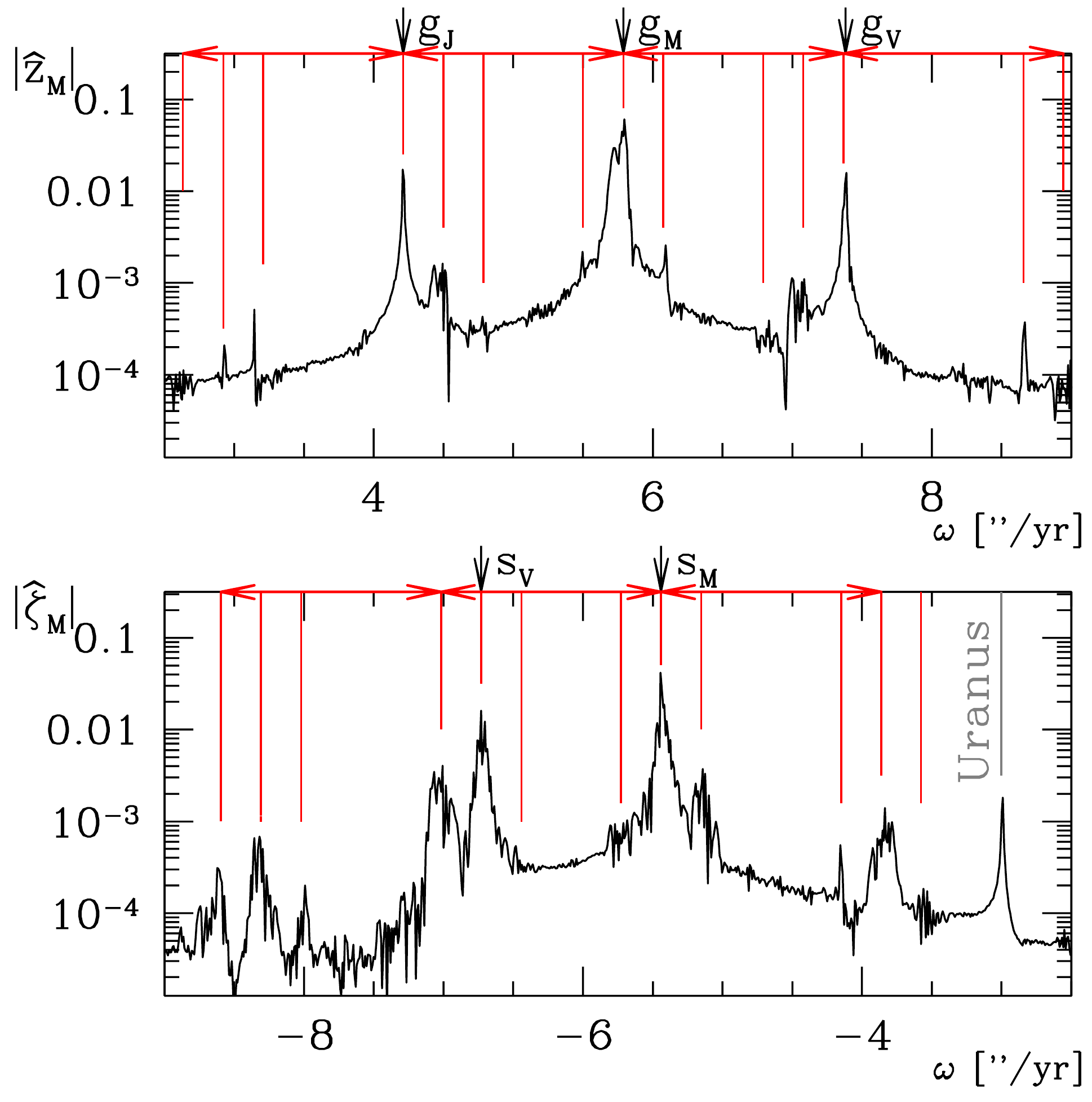}}
\caption{Fourier transform of 
 Mercury's
 $z$ and $\zeta$ (roughly, its
  complex
    eccentricity and inclination) in a SWIFT N-body simulation,
  $\kappa_{\rm nbody}=0.75$: 
  the
  initial $e$'s and $i$'s of all
  planets were pre-multiplied by the factor $\kappa_{\rm nbody}$.  The
  peaks here are broadly similar to those of the $\kappa$-model.  As
  in Fig. \ref{fig:toy75}, they are marked by vertical lines that are
  displaced from $g_M$ and $s_M$ by $m(g_M-g_J)+n(s_M-s_V)$.  The
  agreement between the two figures shows that the $\kappa$-model
  captures much of the physics of the real Mercury.  Nonetheless,
  there are a number of differences.  See text.
\label{fig:f75}
}
\end{figure}

To
compare with the $\kappa$-model, we first consider
cleaner cases by pre-multiplying the 
current
eccentricities and inclinations of all planets by the reduction factor
$\kappa_{\rm nbody}$.  Figure \ref{fig:f75} shows Mercury's Fourier
transforms in a $\kappa_{\rm nbody}=0.75$ integration lasting
$T=150$Myr.  Comparing this with the $\kappa$-model at $\kappa=0.75$
(Fig. \ref{fig:toy75}) shows broad agreement.  In truth, the
parameters for the $\kappa$-model were chosen to match the free and
forced $z$ and $\zeta$ seen in Figure \ref{fig:f75}, i.e. the
frequencies and heights of the four peaks marked $g_J,g_M,s_V,s_M$.
Since there were eight quantities to match, we could do this by
adjusting eight parameters in the $\kappa$-model:
$\gamma,\hat{\gamma},g_J,s_V,\epsilon_J,i_V$, as well as the initial
values of $|z|$ and $|\zeta|$.  Therefore it is not significant that
the forced and free peaks in the two figures agree.  What is
significant is that the other peaks that are generated by nonlinear
couplings of the forced and free peaks also largely agree.  This
indicates that the $\kappa$-model captures much of the nonlinearity as
seen in the real Mercury.

There are, however, at least three
differences of note.  First, the peaks in Figure \ref{fig:f75} are
broader than those in Figure \ref{fig:toy75}.  This is because Figure
\ref{fig:f75} suffers from weak chaos.  But it is remarkable how sharp
the largest peaks are: even though the $e$'s and $i$'s of the Solar
system have only been reduced by 25\%, the resulting chaos is
surprisingly weak.  Note that the integration intervals in the two
figures are the same, $T=4400/\gamma=150$Myr, and hence the finite
width of the peaks in Figure \ref{fig:f75} is not due to the finite
$T$.  A second difference between the two figures is that the
$\hat{z}_M$ peak at frequency $g_M+(g_M-g_J)$ is significantly larger
in the $\kappa_{\rm nbody}$ integration.  That peak is so large
because it is overlapped by a peak forced by Venus's {\it
  eccentricity} mode, which has precession frequency $g_V\approx
2g_M-g_J$.
In other words,    for
$\kappa_{\rm nbody}=0.75$,  Mercury is in a secular resonance
with a librating angle that corresponds  to frequency
$2g_M-g_V-g_J$, and this largely hides the effect
of $g_V$ in the Fourier transform of Figure \ref{fig:f75}.
The third difference between the two figures
is the peak in $\hat{\zeta}_M$ that 
is caused by Uranus's inclination mode.  But this peak appears
to have little dynamical consequence for Mercury.

\begin{figure}
\centerline{\includegraphics[width=0.5\textwidth]{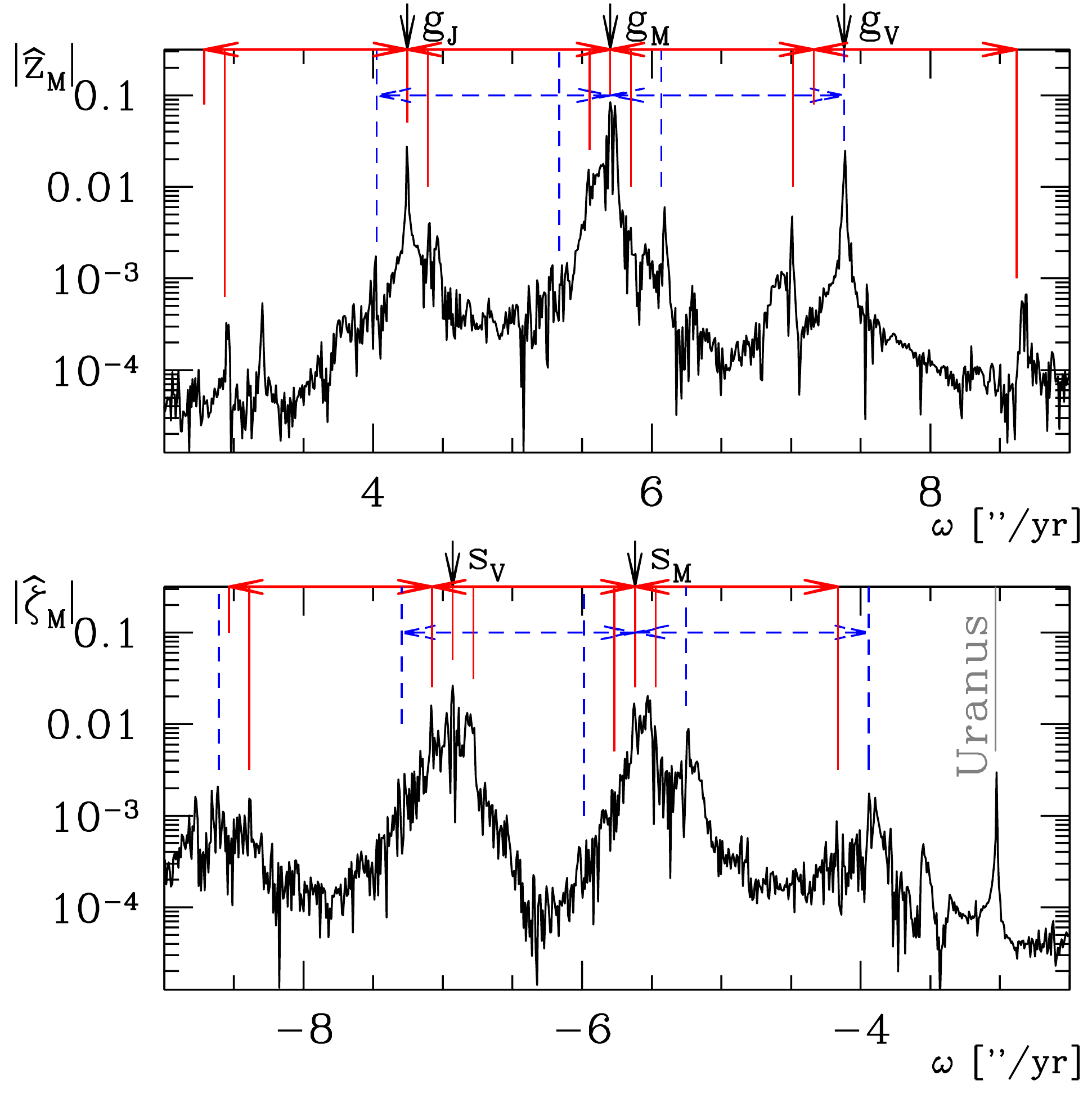}}
\caption{
  Same as Figure \ref{fig:f75}, but with 
 $\kappa_{\rm nbody}$ increased from 0.75 to 0.95.
The motion is more chaotic here, and Venus's eccentricity
forcing at frequency $g_V$ is distinct.  The blue dashed
arrows denote frequency spacings of $g_V-g_M$, and
the blue dashed vertical lines denote peaks due to the $g_V$ mode
and nonlinear couplings generated by that mode.
 \label{fig:f95}
}
\end{figure}

In Figure \ref{fig:f95}, the factor multiplying  the initial $e$'s and $i$'s
has been raised 
to $\kappa_{\rm nbody}=0.95$.  
The resulting motion is more chaotic, as the
widths of the peaks are wider than before, especially
for $\hat{\zeta}_M$. 
 In addition, the frequencies have been 
shifted sufficiently to break Mercury from the
$2g_M-g_V-g_J$ resonance, and the effect of 
Venus's eccentricity forcing is distinct. 
Even though the motion is more chaotic, the principal
forcing peaks and their harmonics are still identifiable.

\begin{figure}
\centerline{\includegraphics[width=0.5\textwidth]{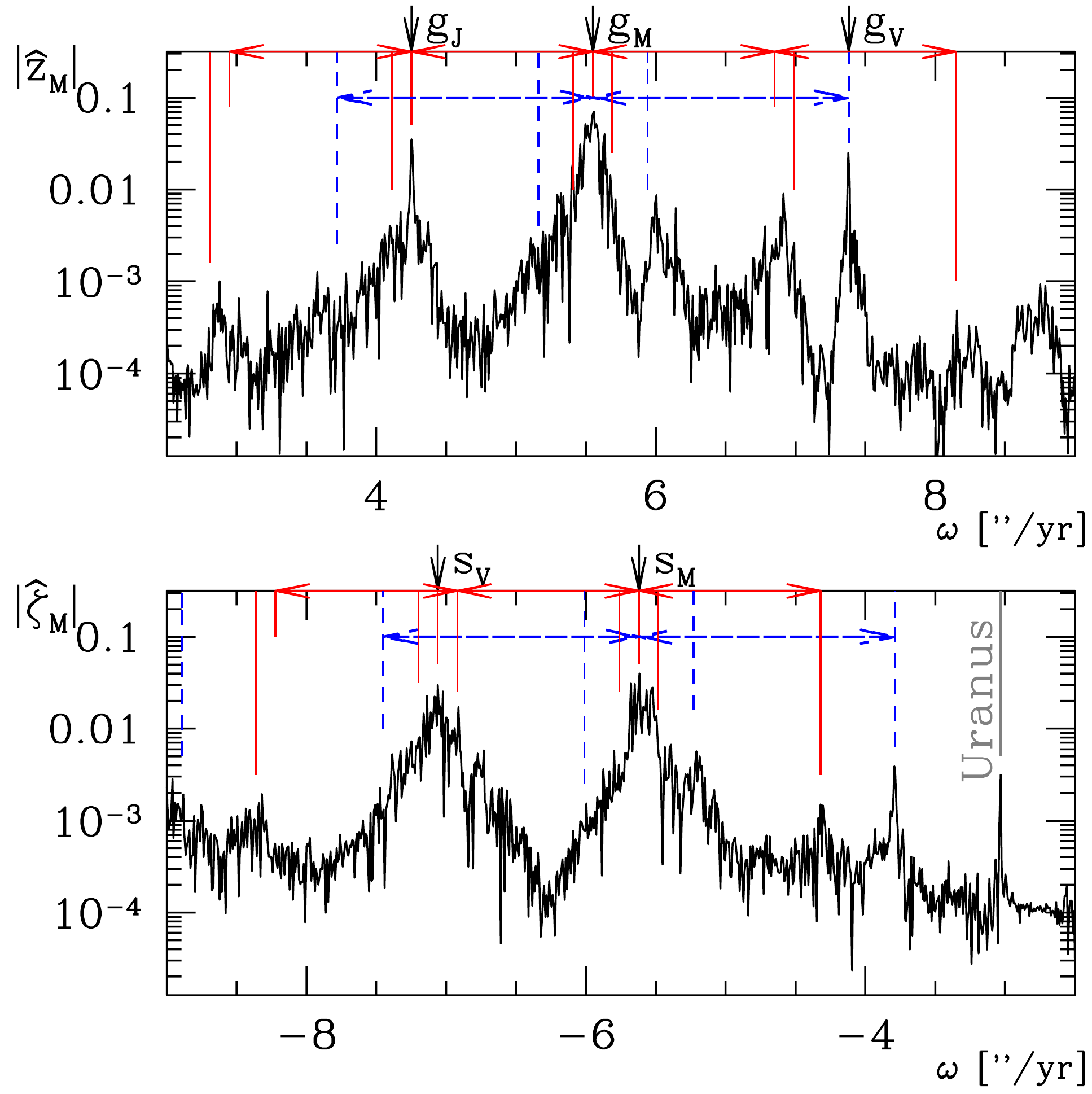}}
\caption{
Same as Figs. \ref{fig:f75}-\ref{fig:f95}, but planets are initialized with 
their true values ($\kappa_{\rm nbody}=1$):
The motion is significantly more chaotic, and
the peaks are less easily identifiable.
Nonetheless, we conclude that the  primary drivers
 of Mercury's
chaos are
Jupiter's eccentricity mode and Venus's inclination mode, 
with
Venus's eccentricity mode playing a supporting role.
\label{fig:fft1}
}
\end{figure}

Figure \ref{fig:fft1} shows the Fourier transform of the real Mercury
($\kappa_{\rm nbody}=1$).  Even though the initial $e$'s and $i$'s
have been increased by only 5\% relative to Figure \ref{fig:f95}, the
motion is significantly more chaotic, and the nonlinearly generated
peaks are less easily identifiable, especially those near Mercury's
free frequencies $g_M$ and $s_M$.  Nonetheless, we conclude from the
progression of Figures \ref{fig:f75}-\ref{fig:fft1}, that the
$\kappa$-model captures much of the physics.  In particular, two
modes---the Jupiter eccentricity mode and the Venus inclination
mode---are primarily responsible for driving Mercury's chaos.  The
most important element lacking from the $\kappa$-model appears
to be the extra forcing by the Venus eccentricity mode.  The fact that
the $\kappa$-model becomes chaotic at
a higher threshold than the real Mercury ($\kappa\sim 1.3$ vs.
$\kappa_{\rm nbody}\sim 1$)
 is likely partly due to that extra forcing
 (i.e. that extra forcing makes the real Mercury more chaotic).
  An additional contributor to the discrepancy between
the two critical $\kappa$'s is that the simple $\kappa$-model does not 
accurately capture nonlinear frequency shifts, while the precise values
of the frequencies are important for where the resonances overlap.
Despite this, the difference between the critical $\kappa$'s is not
large, and this lends support to our claimed origin for Mercury's
chaos.

We note parenthetically that while we only focus on a narrow
range of frequencies in Figure \ref{fig:fft1}, Mercury also
has peaks at $|\omega|\sim 20''$/yr, due to forcing by Earth and
Mars.  However, these peaks have amplitudes $\lesssim 10^{-3}$, 
and appear to have little influence on Mercury's chaotic motion.
Had they been important, one would have expected to see their
influence in Figure \ref{fig:f75}, whereas all the main peaks in that
figure have already been identified.

\subsection{Resonant Angles}

\begin{figure}
\centerline{\includegraphics[width=0.5\textwidth]{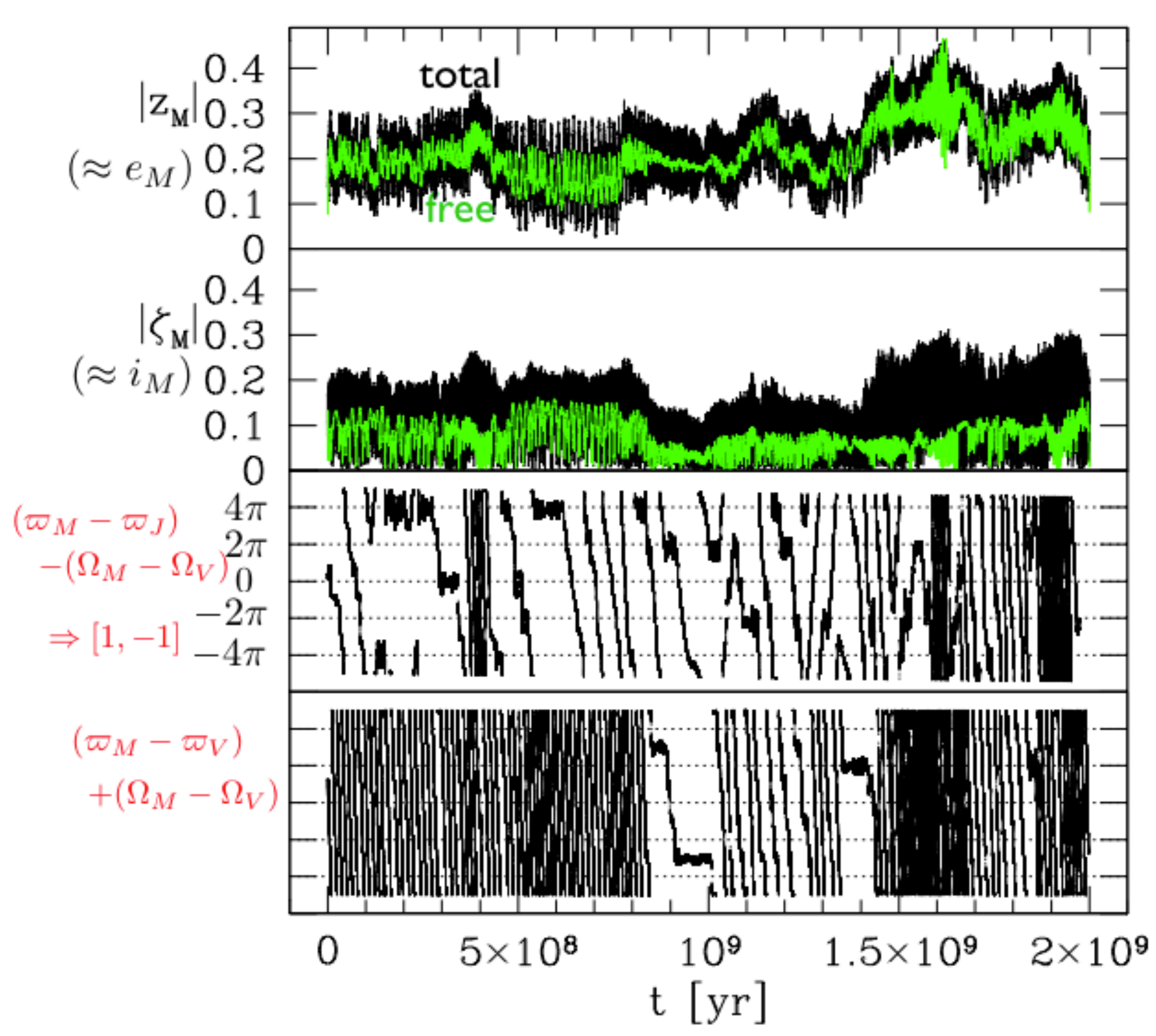}}
\caption{Mercury in an N-body simulation of the Solar system: The top
  panel shows Mercury's $|z_M|$ (approximately its eccentricity) as a
  black curve, for the duration of a 2 Gyr SWIFT simulation.  The
  overplotted green curve is Mercury's free $|z_M|$.  The second panel
  shows the same, but for $\zeta_M$ (approximately its inclination).
  The bottom two panels show the two four-angle combinations involving
  Mercury that were found to undergo libration episodes over the
  course of this simulation.  The plotted angles are the phases of the
  {\it free} orbital elements (see main text).  The angles'
    transitions between libration and circulation are reflected in the
    behavior of of $e_M$ and $i_M$.
\label{fig:tmerc}
}
\end{figure}

Figure \ref{fig:tmerc} shows results from a 2 Gyr SWIFT integration of the 
full Solar system (with no reduction of the initial $e$'s and $i$'s).  
The black curve in the top panel is Mercury's total $|z_M|$, which is
very nearly equal to its total eccentricity (eq. [\ref{eq:zdef}]), and
illustrates the chaotic behavior of Mercury's orbit.  
The overplotted green curve is the absolute value of Mercury's
free $z_M$, which we define to be the part of its total 
$z_M$ that comes from the main peak in Figure \ref{fig:fft1},
i.e., we 
 first take the Fourier
transform of $z_M$, then set to zero all frequencies 
except those satisfying
$4.9''$/yr $<\omega<6.5''$/yr,
and then take 
the inverse Fourier transform.
By plotting the free $z_M$,  the short-term variations
are reduced, and long-term diffusion is clearer.
The second panel in Figure \ref{fig:tmerc} is the same as the
top but for $\zeta_M$; for the free $\zeta_M$, we filter out frequencies outside
of the domain 
$-6.3''$/yr $<\omega<-4.7''$/yr.

The bottom two panels of Figure \ref{fig:tmerc}
 show the two four-angle combinations
involving Mercury
that were found to undergo libration episodes.
The third
 panel shows the angle 
$(\pomega_M-\pomega_J)-(\Omega_M-\Omega_V)$, which 
is the angle that has frequency 
\begin{equation}
\sigma \equiv (g_M - g_J) - (s_M - s_V) \ ,
\label{eq:definesigma}
\end{equation}
i.e., the [1,-1] angle
 (eq. [\ref{eq:sig}]).
(More precisely, we use the angles of the free elements; see below.)
And the bottom panel shows the angle associated
with the frequency 
\be
\sigma'\equiv (g_M-g_V)+(s_M-s_V) \ .
\label{eq:sigmapr}
\ee
\cite{laskar92} has  shown that the  $\sigma$ angle
can change from libration to circulation.  But our finding
that the $\sigma'$ can as well is new.\footnote{
Since the motion is chaotic, it is possible that
$\sigma'$ did not librate at all in the simulations
of \cite{laskar92}  and \cite{SussmanWisdom}.
}
We note that even though both the $\sigma$ and $\sigma'$
angles undergo 
libration episodes, the four-angle combination
that is the sum of the 
two, i.e. the angle associated with $2g_M-g_V-g_J$ does not. 
That angle was found to librate in the $\kappa_{\rm nbody}=0.75$
simulation (Fig. \ref{fig:f75}), and we suspect that it will eventually librate in a long
enough integration of the full Solar system.

The angles displayed in the bottom panels of Figure \ref{fig:tmerc}
were those of the {\it free} elements.  For example, for $\pomega_M$
we first filtered the Fourier transform of $z_M$ as described above,
and took the phase of the free part of $z_M$.  This filtering
procedure is especially important for $\Omega_V$, because Venus's
$\zeta_V$ variations are dominated by forcings due to other modes,
including the Mercury-, Earth-, and Mars-dominated modes
\citep{laskar90}, whereas we wish $\Omega_V$ to denote the phase of
the Venus-dominated mode.  Therefore, we first filter the Fourier
transform of $\zeta_V$, keeping only frequencies $-7.7''$/yr
$<\omega<-6.3''$/yr, and use for $\Omega_V$ the phase of the filtered
$\zeta_V$.  Similarly, we filter $z_M,\zeta_M, z_V$, and $z_J$ with
appropriate windows to obtain the other angles of interest.  Our
method of filtering for extracting mode angles differs from that of
\cite{laskar90}, who extracts mode angles by projecting onto the
numerically determined nonlinear ``proper modes.''  We have
experimented with a number of different methods, and also with
changing the size of the filter window, and the duration of the
integration, and found that our filtering method 
   is simple to
  implement, is computationally efficient, and gives reliable results.

\begin{figure}
\centerline{\includegraphics[width=0.5\textwidth]{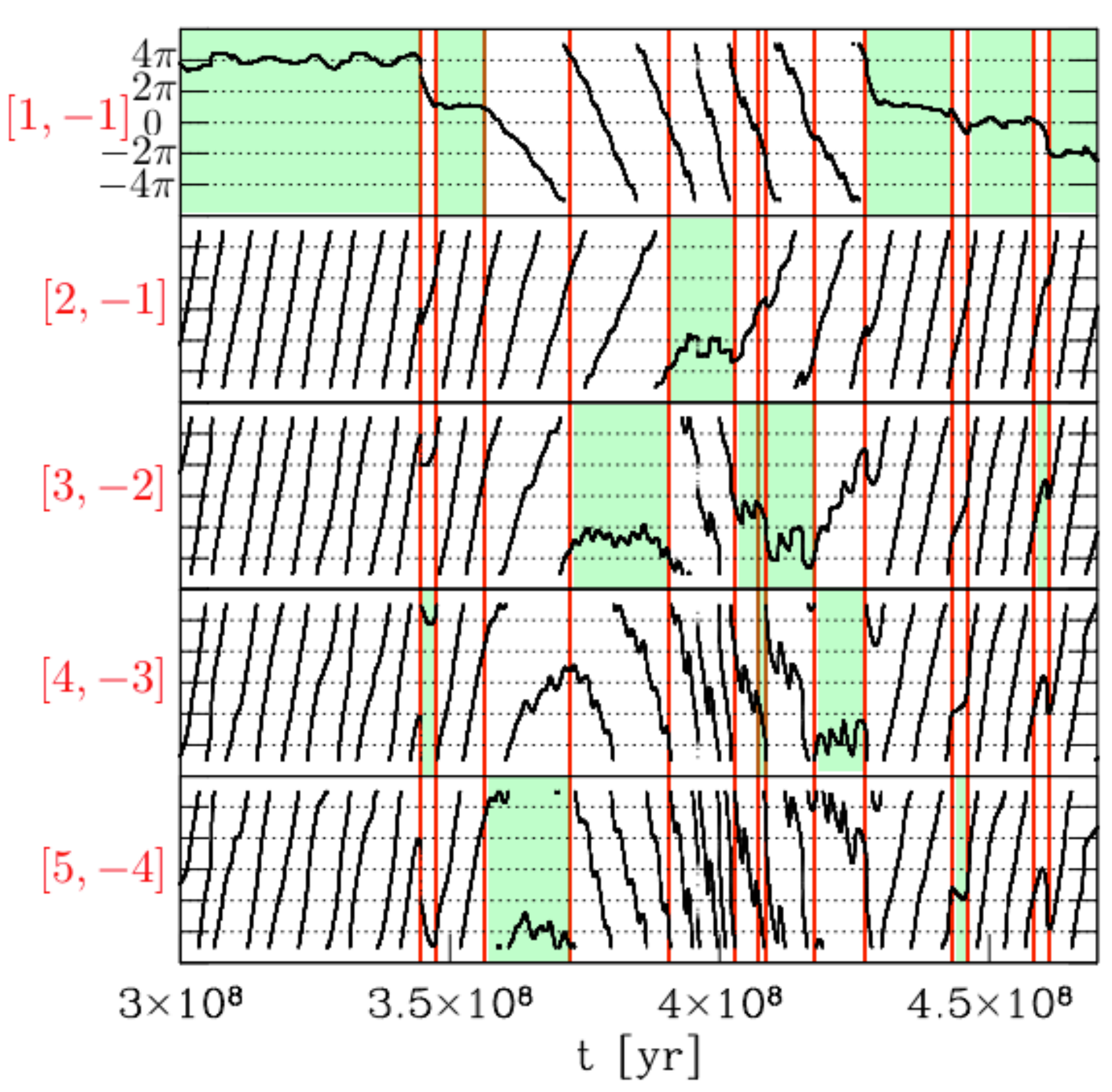}}
\caption{Chain of librating angles for the real Mercury:
This is the same N-body simulation as in Figure \ref{fig:tmerc}, 
focusing
on the time when the [1,-1] angle transitions to its
first extended period of circulation and then back to libration.
Each panel shows the angle that has frequency 
$[m,n]=m(g_M-g_J)+n(s_M-s_V)$, for various
values of $[m,n]$.  The angles 
alternately librate, showing that the
chaos is at least partly caused by the overlap of these
resonances, as in the
  $\kappa$-model
(Fig.  \ref{fig:resang}).
The plotted angles
are the phases of the free orbital elements.
\label{fig:rm}
}
\end{figure}

In addition to the two four-angle
combinations of Figure \ref{fig:tmerc}, one might suspect that
 there are
many more higher-order combinations that librate when both 
of those angles simultaneously circulate, as in the Hamiltonian
model.  
In Figure \ref{fig:rm}, we zoom into the episode when the $[1,-1]$
angle first undergoes an extended period of circulation, and
plot some higher order combinations associated with the frequencies
$[m,n]\equiv m(g_M-g_J)+n(s_M-s_V)$.
It can be seen that these angles librate in turn, just
as in the Hamiltonian model (compare with Fig.  \ref{fig:resang}).
  This provides another demonstration
that the physics of the Hamiltonian model is similar to that
of the real Mercury.

\section{Summary and Discussion}

\label{sec:summary}

We have shown how secular chaos is driven by the
overlap of secular resonances, both for a test particle
modelled with a simplified Hamiltonian, 
and for the real Mercury.
To linear order, secular frequencies are constant.
But 
nonlinearities
can shift planets into and out of secular resonance with 
each other, and when two resonances overlap, chaos 
results.

In \S\S \ref{sec:sem}-\ref{sec:iej}, 
we focused on the evolution of a test particle in 
the presence of multiple massive planets.
The test particle was evolved to 
leading nonlinear order, and 
the
$e$'s, $i$'s, and precession rates of the planets
(or more properly of the planet modes)
 were taken to be constant.
We first considered the  simple case with zero 
inclinations, as was first worked out by \cite{sid90}.
In \S \ref{sec:iej}, we generalized to 
 non-zero inclinations, 
when
the test particle
 comes
under the influence of 
 one eccentric and one inclined
planet mode.  In that case, the particle has two
free frequencies, its apsidal and nodal frequencies
($g$ and $s$).  Each of these is altered
 by the particle's $e$ and $i$.
Therefore each resonance traces out a one-dimensional
curve in the particle's $e$-$i$ plane, 
or equivalently in its $p_e$-$p_i$ plane.
A simple way to map out the 
dynamics is with the ``mean momentum map''
(MMM),
whereby the particle's time-averaged
$p_e$ and $p_i$ are plotted 
against each other for different initial conditions.
This shows
 where the resonances are, how wide they 
are, and how their overlap leads to chaos (Figs. \ref{fig:low}-\ref{fig:pphi}).
We calculated analytically the locations and widths of the four strongest  
resonances---the [1,0], [0,1], and [1,$\pm 1$]---and showed that
these agreed with the numerical MMM results (Figs. \ref{fig:pplot}-\ref{fig:pplothi}).
Chaos in this case
emerges from the overlap of resonances of the form $[m,n]$ (eq. [\ref{eq:sig}]), 
with typically $n=m\pm 1$ and $m$ a small integer.
This may be seen in the MMM, in
 surfaces of section (Fig. \ref{fig:sos}), and also by explicitly
tracing the chain of librating angles
(Fig. \ref{fig:resang}).
We also examined the
test particle's trajectories in Fourier space
(Figs. \ref{fig:toy75}-\ref{fig:toyfftall}).

In \S \ref{sec:realmercury}, we considered the orbital evolution
of Mercury in N-body simulations.  We showed that despite
all the simplifications we made in the Hamiltonian models, 
the real Mercury behaved in a qualitatively similar manner.
In particular:
\bi
\item Mercury's chaos is primarily driven by the $s_V$ and $g_J$ modes
  (i.e. the Venus- and Jupiter-dominated $i$ and $e$ modes), although
  the $g_V$ mode also plays a role.  The nonlinear couplings between
  those modes and Mercury's own free modes (with frequencies $g_M$ and
  $s_M$) are primarily responsible for Mercury's chaos
  (Figs. \ref{fig:f75}-\ref{fig:fft1}).
\item There are a slew of resonant angles that drive Mercury's chaos.
  Just as in the Hamiltonian model, a chain of resonant angles of the
  form $m(\pomega_M-\pomega_J)+n(\Omega_M-\Omega_V)$ show sequential
  librations, for integers $[m,n]$ (Fig. \ref{fig:rm}), where the
  angles refer to the phases of the free orbital elements.  In
  addition, we identified a new four-angle combination,
  $(\pomega_M-\pomega_V)+(\Omega_M-\Omega_V)$, that can also undergo
  libration episodes (Fig. \ref{fig:tmerc}).
\item Mercury is perched on the threshold of chaos.  If one reduces
  the $e$'s and $i$'s of the planets by only 25\%, Mercury's motion
  becomes nearly regular (Fig. \ref{fig:f75}).  This behavior is
    also apparent in the Hamiltonian model (Fig. \ref{fig:pphi}).  We
    have also performed a $\kappa_{\rm nbody}=1.2$ simulation, in
    which the planets' initial $e$'s and $i$'s were increased $20\%$
    (not shown).  The result was violent instability, with Mercury
    ejected in $\sim 100$ Myr.  
\ei

Having identified the secular resonances responsible for Mercury's
  chaos, and calculated the widths and locations of those resonances,
  we can calculate the threshold for Mercury's chaos.  We do that here
  in an approximate way.  First, since Mercury's apsidal and nodal
  frequencies differ from $g_J$ and $s_V$ by $\sim 20\%$, the
  co-ordinates in the $p_e$-$p_i$ plane where the two resonances
  overlap are around half that, or $p_{e**}\sim p_{i**}\sim 0.1$
  (eq. [\ref{eq:pess}]).  Second, the width of the chaotic overlap
  zone between the [1,-1] and the [1,0] (or [0,1]) is $\sim 2
  (\epsilon_Ji_J)^{1/4}(p_{e**}p_{i**})^{1/8}$ (eq. [\ref{eq:over}];
   we include here an extra factor of 2 to account
  for the difference between the half- and full-width, as described
  in \S \ref{sec:theory})  Therefore if $\epsilon_J\sim i_J\gtrsim p_{e**}^{3/2}/4\sim
  0.01$, then the region of chaos will encroach upon the origin of the
  $p_e$-$p_i$ plane.  This explains why Mercury can be chaotic even
  though the eccentricities and inclinations in the Solar system are
  at the level of  a few percent.

The work discussed in this paper can be extended in a number of
directions.  The theory can be extended to order-unity eccentricities
and inclinations.  Although that case will be more complicated, we
suspect that the basic structure will remain, with resonant zones in
the $e$-$i$ plane whose overlap leads to chaos.  One can also attempt
to build a theory that includes long-term diffusion and massive
planets, as well as incorporating MMR's.

A number of applications also come to mind, such as quantifying
Mercury's chaotic diffusion and understanding how it came about that
Mercury is perched on the threshold of chaos.  The latter seems to be
a clue for understanding how the Solar system arrived at its current
marginally stable state.  
It would also be interesting
  to investigate the role of $\sigma'$ (eq. [\ref{eq:sigmapr}]). Is it an
  unnecessary coincidence for Mercury's chaos?
  
Our theory can also be applied to  
 Earth and
Mars, for which librating angles have been identified that are similar
to those we found for Mercury, i.e.  angles of the form
$m(\pomega_{\rm mars}-\pomega_{\rm earth})+ n(\Omega_{\rm
  mars}-\Omega_{\rm earth})$, with $[m,n]=[1,-1], [2,-1],$ and
$[3,-2]$ \citep{laskar92,SussmanWisdom}. 
We also propose
  that secular chaos can play a role in shaping extra-solar planetary
systems \citep{paper2}, and hence the theory of secular
chaos might be applicable to extra-solar planets as well.

\appendix
\section{\bf Appendix A: Fourth Order Secular Hamiltonian}
\label{sec:app1}

In this Appendix, we give the expression for the secular Hamiltonian of a test
particle perturbed by an external planet, where both particle
and planet are orbiting a star.  The Hamiltonian 
is expanded to fourth order in the particle's eccentricity and inclination, and
to leading order in the ratio of semi-major axes.
The energy per unit mass of the test particle is
\be
E = -{GM_\odot\over 2a}-{Gm'\over  a'}R  \ , 
\label{eq:eperunit}
\ee
where  $M_\odot$ is the mass of the star, $a$ and $a'$ are,
respectively, the test particle's and planet's semimajor axes, $m'$
is the planet's mass,
and $R$ is the disturbing
function.
We approximate $R$ by
only retaining the secular terms up to fourth order in
$e$ and $s\equiv\sin(i/2)$ and second order in $\alpha\equiv a/a'$
(except for the $f_{10}$ term whose leading contribution is $O(\alpha^3 e^2)$):
\beqn
R &\approx &f_2 e^2 + f_3  s^2
+ f_5 e^2 e'^2 
+ f_7(e^2s^2+e^2s'^2+e'^2s^2)+f_8 s^4 
+f_9s^2s'^2 + f_{10} e e'\cos(\pomega-\pomega')  \nonumber \\
&+&  \left(f_{14}ss' + f_{15}ss'(e^2+e'^2)+f_{16}ss'(s^2+ s'^2)\right)\cos(\Omega-\Omega')
+f_{18}e^2s^2\cos(2\pomega-2\Omega) 
+ f_{21}e^2ss'\cos({2\pomega-\Omega'-\Omega})
 \nonumber \\
 &+&
 f_{18}e^2s'^2\cos(2\pomega-2\Omega')
+
f_{26} s^2 s'^2 \cos(2\Omega-2\Omega')  \ ,
\eeqn
in the notation of the appendix of \cite{MD00}.
The $f_i$ are functions of $\alpha$ that may be expressed as
 sums of Laplace coefficients and their derivatives.
We drop terms that are independent of the test particle's orbital elements.

In this paper, we work with a scaled Hamiltonian, 
$H\equiv -2E/\sqrt{GM_\odot a}$
(eq. [\ref{eq:hamscal}]), and hence
\be
H=\gamma{8R\over 3\alpha^2} \ ,
\label{eq:hdef}
\ee
dropping the Keplerian term in $E$ because it is irrelevant for secular
dynamics, and defining
\be
\gamma\equiv {3\over 4}{m'\over M_\odot}\alpha^3\left(GM_\odot\over a^3  \right)^{1/2} \ ,
\label{eq:gammadef} 
\ee
which is the test particle's secular free precession frequency based on linear theory.
The scaled disturbing function
$8R/(3\alpha^2)$
is a sum of terms that are listed
in Table \ref{tab:simp}, after expanding the $f_i$ to $O(\alpha^2)$, 
and $f_{10}$ to $O(\alpha^3)$.

\begin{table}[t]
\caption{Terms in scaled disturbing function ${8R\over 3\alpha^2}
 =
  \sum_i c_ih_i$.
  The Hamiltonian is $H=\gamma \sum_i c_ih_i$.
  The variables
  $z$ and $\zeta$ (defined in eqs. [\ref{eq:zdef}]-[\ref{eq:zetadef}])
   are approximately the complex eccentricity and inclination.
    Terms 1-4 are z-only.
 Terms 11-17 are $\zeta$ only.  Terms 21+ are mixed. c.c. denotes
 complex conjugate.}

\begin{minipage}{160mm}
\begin{tabular}{|c|cccc|ccccccc|}
\hline
$i$&1&2&3&4&11&12&13&14&15&16&17
 \\ \hline\hline
$c_i$&$1$&  $-{5\over 4}\alpha$ &   ${3\over 2}$ &  $-{1\over 4}$ 
&$-1$&$1$&${3\over 2}$&${1\over 4}$&$-{5\over 8}$&$-{5\over 8}$&${1\over4}$
 \\
$h_i$&  $|z|^2$  &$ z^*z' + {\rm c.c.} $
&$|z|^2|z'|^2$
&$|z|^4$
&$|\zeta|^2$
&$\zeta^*\zeta' + {\rm c.c.}$
&$|\zeta|^2|\zeta'|^2$
&$|\zeta|^4$
&$|\zeta|^2\zeta^*\zeta'+{\rm c.c.}$
&$|\zeta'|^2\zeta^*\zeta'+{\rm c.c.}$
&$\zeta^{*2}\zeta^{'2}+{\rm c.c.}$\\ \hline
\end{tabular}
\end{minipage}
\begin{minipage}{160mm}
\begin{tabular}{|c|cccccccc|}
\hline
$i$&21&22&23&24&25&26&27&28 \\ \hline
$c_i$
&$-2$
&$-{3\over 2}$
&$-{3\over 2}$
&${7\over 4}$
&${7\over 4}$
&${5\over 4}$
&${5\over 4}$
&$-{5\over 2}$
\\
$h_i$
&$|z|^2|\zeta|^2$
&$|z'|^2|\zeta|^2$
&$|z|^2|\zeta'|^2$
&$|z|^2\zeta^*\zeta'+{\rm c.c.}$
&$|z'|^2\zeta^*\zeta'+{\rm c.c.}$
&$z^{*2}\zeta^2 + {\rm c.c.}$
&$z^{*2}\zeta'^{2} + {\rm c.c.}$
&$z^{*2}\zeta\zeta'+{\rm c.c.}$
\\ \hline
\end{tabular}
\end{minipage}
\label{tab:simp}
\end{table}

\section{\bf Appendix B: Width of the [1, -1] and [1, 1] resonances from von Zeipel
Transformation}

We start from Hamiltonian (\ref{eq:hamrot}), which we reproduce
here as
\be
H(p_e,q_e;p_i,q_i)=-{p_e^2-p_i^2\over 4} 
+\Delta p_e +\Delta_sp_i-2p_ep_i
-2\epsilon_J\sqrt{p_e}\cos q_e
+2i_J\sqrt{p_i}\cos q_i \ , 
\ee
setting $\gamma=1$, $q_e\equiv \pomega_-=\pomega-g_Jt$, and 
$q_i\equiv \Omega_-=\Omega-s_Jt$.
We solve this Hamiltonian perturbatively, treating $\epsilon_J$ and
$i_J$ as the small parameters.  This is equivalent to
expanding in the test particle's forced eccentricity and inclination, 
assumed to be much smaller than the free $e$ and $i$.
We transform to capitalized variables with the von Zeipel generating function
\be
F(P_e,q_e;P_i,q_i)=P_eq_e+P_iq_i+k_e(P_e,P_i)\sin q_e+k_i(P_e,P_i)\sin q_i \ ,
\ee
where the first two terms generate the identity transformation, and 
the functions $k_e$ and $k_i$ are first order in $\epsilon_J$
and $i_J$; their form will
 be chosen to ``kill'' the cosine terms
in the Hamiltonian to leading order.
The von Zeipel generating function transforms variables as follows
\beqn
p_e&=&P_e+k_e\cos q_e \\
p_i&=&P_i+k_i\cos q_i \\
Q_e&=&q_e+\partial_{P_e}k_e\sin q_e \\
\ Q_i&=&q_i+\partial_{P_i}k_i\sin q_i \ .
\eeqn
Inserting into the Hamiltonian yields
\be
H(P_e,Q_e;P_i,Q_i)=-{P_e^2-P_i^2\over 4}+\Delta P_e+\Delta_s P_i-2P_eP_i
-k_ek_i\cos (Q_e- Q_i)-k_ek_i\cos(Q_e+Q_i) 
\ee
to second order, 
after setting
\beqn
k_e&=&{2\epsilon_J\sqrt{P_e}\over \Delta-P_e/2-2P_i} \\
k_i&=&{-2i_J\sqrt{P_i}\over \Delta_s+P_i/2-2P_e}  \ ,
\eeqn
to eliminate the first order terms.
The two cosine terms in this Hamiltonian are the [1, -1] and
[1, 1] resonances, respectively (see eqs. [\ref{eq:freq1}]-[\ref{eq:freq2}] and following).
Note that we have dropped 
second order terms in the Hamiltonian that 
are proportional to $\cos^2Q_e$, $\sin^2Q_e$, $\cos^2Q_i$, 
and $\sin^2Q_i$, because these have little effect on the [1, 1] and [1, -1]
resonances; but they generate new frequency components which 
are important for higher order resonances.  

 To leading order, $k_e$
is twice the product of the free eccentricity $(\sqrt{P_e})$ with the forced
eccentricity, where the forced eccentricity differs from the linear
expression ($\epsilon_J/\Delta$;  see eq. [\ref{eq:linsol}])
by the terms $P_e/2+2P_i$ in the denominator,  which arise from the nonlinear shift
of the frequency (eqs. [\ref{eq:freq1}]-[\ref{eq:freq2}]).  Similarly $k_i$ is twice the product of the
free and forced inclinations.  Hence the strengths of the [1, -1] and
[1, 1] resonances are proportional to the products of the free and forced
eccentricities and inclinations, as argued qualitatively in 
\S \ref{sec:iej}.

To determine the width of the [1, -1] resonance, we drop the
last cosine
term in the above Hamiltonian.
Since
$P_+\equiv P_e+P_i$ is an integral of the motion, 
we may re-write the Hamiltonian as
$H(P_e,Q_-)=2(P_e-P_*)^2-k_ek_i\cos Q_-$, 
dropping a constant and defining $Q_-\equiv Q_e-Q_i$
and $P_*\equiv (5/8)P_+-(1/4)(\Delta_g-\Delta_s)$.
Therefore the half-width of the resonance is
\be
\delta P_e=\sqrt{|k_ek_i|} \ . \label{eq:widapp}
\ee
 We take the amplitude of the cosine term to be fixed at its value
at resonance center  \citep[e.g.,][]{Chirikov}.  Since $P_e+P_i$ is constant, the
half-width in $P_i$ is the same, $\delta P_i=\sqrt{|k_ek_i|}$.


\bibliographystyle{apj}
\bibliography{ms}

\begin{thebibliography}{23}
\expandafter\ifx\csname natexlab\endcsname\relax\def\natexlab#1{#1}\fi

\bibitem[{{Batygin} \& {Laughlin}(2008)}]{Batygin}
{Batygin}, K. \& {Laughlin}, G. 2008, \apj, 683, 1207

\bibitem[{{Chirikov}(1979)}]{Chirikov}
{Chirikov}, B.~V. 1979, \physrep, 52, 263

\bibitem[{Henrard \& Lemaitre(1983)}]{Henrard83}
Henrard, J. \& Lemaitre, A. 1983, Celestial Mechanics (ISSN 0008-8714), 30, 197

\bibitem[{{Laskar}(1989)}]{Laskar89}
{Laskar}, J. 1989, \nat, 338, 237

\bibitem[{{Laskar}(1990)}]{laskar90}
---. 1990, Icarus, 88, 266

\bibitem[{{Laskar}(1992)}]{laskar92}
{Laskar}, J. 1992, in IAU Symposium, Vol. 152, Chaos, Resonance, and Collective
  Dynamical Phenomena in the Solar System, ed. {S.~Ferraz-Mello}, 1--+

\bibitem[{{Laskar}(1993)}]{laskar93}
---. 1993, Celestial Mechanics and Dynamical Astronomy, 56, 191

\bibitem[{{Laskar}(1996)}]{Laskar96}
---. 1996, Celestial Mechanics and Dynamical Astronomy, 64, 115

\bibitem[{{Laskar}(2008)}]{Laskar08}
---. 2008, Icarus, 196, 1

\bibitem[{{Laskar} \& {Gastineau}(2009)}]{Laskar09}
{Laskar}, J. \& {Gastineau}, M. 2009, \nat, 459, 817

\bibitem[{{Lecar} {et~al.}(2001){Lecar}, {Franklin}, {Holman}, \&
  {Murray}}]{Lecar}
{Lecar}, M., {Franklin}, F.~A., {Holman}, M.~J., \& {Murray}, N.~J. 2001,
  \araa, 39, 581

\bibitem[{{Levison} \& {Duncan}(1994)}]{levisonduncan}
{Levison}, H.~F. \& {Duncan}, M.~J. 1994, Icarus, 108, 18

\bibitem[{{Murray} \& {Dermott}(2000)}]{MD00}
{Murray}, C.~D. \& {Dermott}, S.~F. 2000, {Solar System Dynamics} (Cambridge
  University Press)

\bibitem[{{Murray} \& {Holman}(1999)}]{murrayholman99}
{Murray}, N. \& {Holman}, M. 1999, Science, 283, 1877

\bibitem[{{Naoz} {et~al.}(2010){Naoz}, {Farr}, {Lithwick}, {Rasio}, \&
  {Teyssandier}}]{smadar}
{Naoz}, S., {Farr}, W.~M., {Lithwick}, Y., {Rasio}, F.~A., \& {Teyssandier}, J.
  2010, ArXiv e-prints

\bibitem[{{Ogilvie}(2007)}]{Ogilvie07}
{Ogilvie}, G.~I. 2007, \mnras, 374, 131

\bibitem[{{Quinn} {et~al.}(1991){Quinn}, {Tremaine}, \& {Duncan}}]{Quinn91}
{Quinn}, T.~R., {Tremaine}, S., \& {Duncan}, M. 1991, \aj, 101, 2287

\bibitem[{{Sidlichovsky}(1990)}]{sid90}
{Sidlichovsky}, M. 1990, Celestial Mechanics and Dynamical Astronomy, 49, 177

\bibitem[{{Sussman} \& {Wisdom}(1988)}]{SussmanWisdom88}
{Sussman}, G.~J. \& {Wisdom}, J. 1988, Science, 241, 433

\bibitem[{{Sussman} \& {Wisdom}(1992)}]{SussmanWisdom}
---. 1992, Science, 257, 56

\bibitem[{{Wisdom}(1983)}]{wisdom83}
{Wisdom}, J. 1983, Icarus, 56, 51

\bibitem[{{Wisdom} \& {Holman}(1991)}]{WisdomHolman}
{Wisdom}, J. \& {Holman}, M. 1991, \aj, 102, 1528

\bibitem[{{Wu} \& {Lithwick}(2010)}]{paper2}
{Wu}, Y. \& {Lithwick}, Y. 2010, ArXiv e-prints

\end{thebibliography}

\end{document}